\documentclass[%
 reprint,
 amsmath,amssymb,
 aps,
]{revtex4-2}

\usepackage{graphicx}
\usepackage{dcolumn}
\usepackage{bm}
\usepackage{color,soul}
\usepackage{amsmath}
\usepackage[utf8]{inputenc}
\usepackage[english]{babel}
\usepackage{setspace}

\usepackage{chngcntr}
\counterwithout{figure}{section}
\counterwithout{equation}{section}

\begin{document}


\title{Enhanced Meta-Displays Using Advanced Phase-Change Materials}

\author{Omid Hemmatyar$^{1,\dagger}$}
\author{Sajjad Abdollahramezani$^{1,\dagger}$}
\author{Ioannis Zeimpekis$^{2}$}
\author{Sergey Lepeshov$^{3}$}
\author{Alex Krasnok$^{4}$}
\author{Asir Intisar Khan$^{5}$}
\author{Kathryn M. Neilson$^{5}$}
\author{Christian Teichrib$^{6}$}
\author{Tyler Brown$^{1}$}
\author{Eric Pop$^{5}$}
\author{Daniel W. Hewak$^{2}$}
\author{Matthias Wuttig$^{6}$}
\author{Andrea Al{\`u}$^{4, 7}$}
\author{Otto L. Muskens$^{8}$}
\author{Ali Adibi$^{1}$}


\affiliation{$^{1}$School of Electrical and Computer Engineering, Georgia Institute of Technology, 778 Atlantic Drive NW, Atlanta, Georgia 30332-0250, US}
\affiliation{$^{2}$Zepler Institute, Faculty of Engineering and Physical Sciences, University of Southampton,
SO17 1BJ Southampton, United Kingdom}
\affiliation{$^{3}$ITMO University, St. Petersburg 197101, Russia}
\affiliation{$^{4}$Photonics Initiative, Advanced Science Research Center, City University of New York, New York, New York 10031, United States}
\affiliation{$^{5}$Department of Electrical Engineering, Department of Materials Science and Engineering, Precourt Institute for Energy, Stanford University, Stanford, California 94305, United States}
\affiliation{$^{6}$Physikalisches Institut IA, RWTH Aachen, Sommerfeldstrasse 14, 52074 Aachen, Germany}
\affiliation{$^{7}$Physics Program, Graduate Center, City University of New York, New York, New York 10016, United States}
\affiliation{$^{8}$Physics and Astronomy, Faculty of Engineering and Physical Sciences, University of Southampton, SO17 1BJ Southampton, United Kingdom,}
\affiliation{$^{\dagger}$These authors contributed equally to this work.}

\date{\today}

\begin{abstract}

Structural colors generated due to light scattering from static all-dielectric metasurfaces have successfully enabled high-resolution, high-saturation, and wide-gamut color printing applications. Despite recent advances, most demonstrations of these structure-dependent colors lack post-fabrication tunability. This hinders their applicability for front-end dynamic display technologies. Phase-change materials (PCMs), with significant contrast of their optical properties between their amorphous and crystalline states, have demonstrated promising potentials in reconfigurable nanophotonics. Herein, we leverage tunable all-dielectric reflective metasurfaces made of newly emerged classes of low-loss optical PCMs, i.e., antimony trisulphide (Sb$_2$S$_3$) and antimony triselenide (Sb$_2$Se$_3$), with superb characteristics to realize switchable, high-saturation, high-efficiency and high-resolution dynamic meta-pixels. Exploiting polarization-sensitive building blocks, the presented meta-pixel can generate two different colors when illuminated by either one of two orthogonally polarized incident beams. Such degrees of freedom (i.e., material phase and polarization state) enable a single reconfigurable metasurface with fixed geometrical parameters to generate four distinct wide-gamut colors. We experimentally demonstrate, for the first time, an electrically-driven micro-scale display through the integration of phase-change metasurfaces with an on-chip heater formed by transparent conductive oxide. Our experimental findings enable a versatile platform suitable for a wide range of applications, including tunable full-color printing, enhanced dynamic displays, information encryption, and anti-counterfeiting.

\end{abstract}


\maketitle


\section*{Introduction}

In the past decades, absorption and emission of light from organic dyes and chemical pigments have been the most common color generation mechanisms in color-imaging and display devices \cite{daqiqeh2020nanophotonic}. Nevertheless, there are still several challenges with the developed technologies, such as environmental hazards, vulnerability to high-intensity light, and limited scalability to smaller pixel sizes. In order to address these issues, structural colors have emerged as compelling alternatives. Structural colors are observed in numerous natural species, whose bright features arise from light scattering and interference in micro/nanostructured patterns of their skins or scales \cite{vukusic1999quantified}. Inspired by nature and enabled by recent advancement in nanofabrication, artificial structural colors generated via a resonant interaction between incident white light and miniaturized building blocks in optical metasurfaces \cite{yu2011light,krasnok2012all, decker2015high,kuznetsov2016optically}, i.e., arrays of subwavelength patterned nanostructures, have gained great attention in recent years. In this context, plasmonic metasurfaces made of gold, silver and aluminum nanostructures have been extensively used to generate structural colors based on plasmon resonances \cite{duan2017dynamic}. Despite their versatility, the broad and weak plasmon resonances, imposed by the significant inherent ohmic loss of the constituent metallic materials, result in low color saturation and purity \cite{kristensen2016plasmonic}.

\begin{figure*}[htbp]
\centering
\includegraphics[width=1\linewidth, trim={0cm 0cm 0cm 0cm},clip]{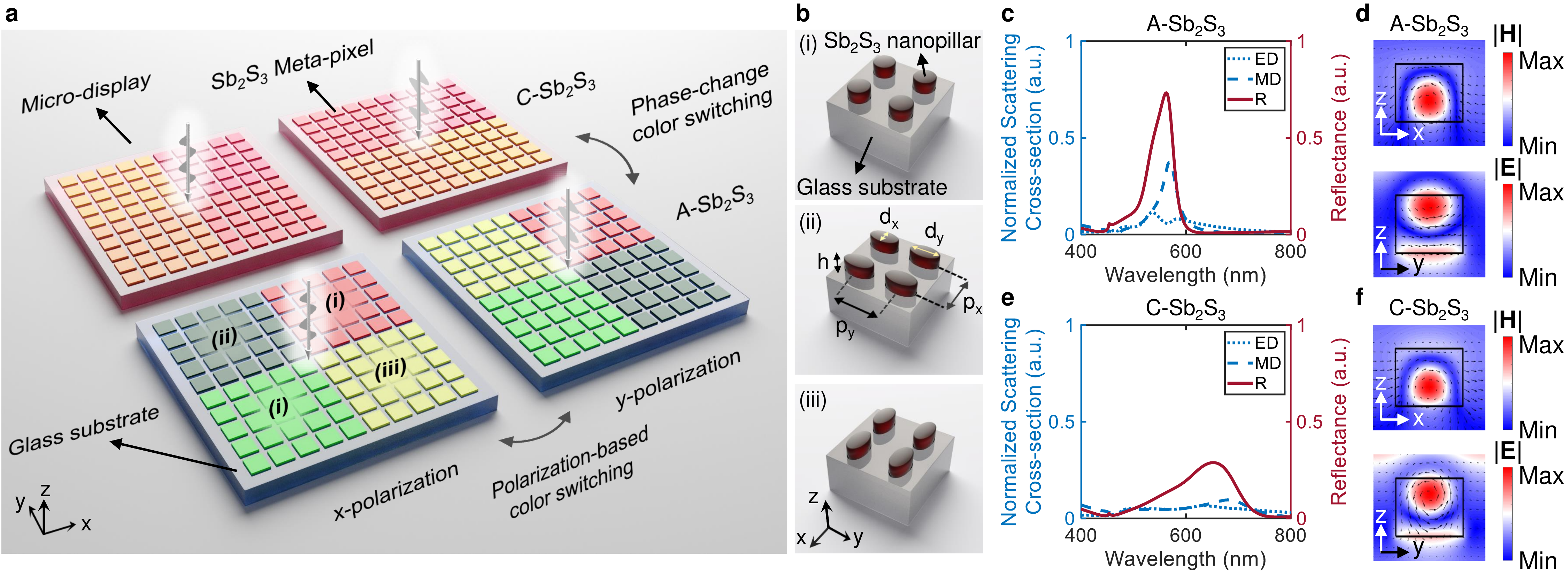}
\caption{\textbf{Working principle of a polarization-encoded dynamic display composed of phase-change meta-pixels}. \textbf{a},\textbf{b}, Schematic representation of a reflective display consisting of phase-change meta-pixels. Each meta-pixel is a metasurface formed by a periodic arrangement of Sb$_2$S$_3$ nanopillars shown in (\textbf{b}), which can generate four different colors; two colors for each polarization attributed to the amorphous and crystalline phases of Sb$_2$S$_3$ (i.e. A-Sb$_2$S$_3$ and A-Sb$_2$S$_3$, respectively) nanopillars or two colors for each Sb$_2$S$_3$ phase (corresponding to x-polarized and y-polarized incident white light). For all metasurfaces, the height ($h$) of the nanopillars is fixed while their periodicity in x- and y-directions (i.e., $p_x$ and $p_y$, respectively) change to generate different colors. The major and minor axes of the nanopillars in x- and y-directions are proportional to the corresponding periodicities in those directions with a constant aspect ratio, i.e. $d_{x,y} = \alpha \, p_{x,y}$, in which $\alpha$ is constant. The colors shown in (\textbf{a}) correspond to Sb$_2$S$_3$ metasurfaces with (i) $p_x=p_y=310$ nm (for green) and $p_x=p_y=390$ nm (for red), (ii) $p_x=310$ nm, $p_y=390$ nm, and (iii) $p_x=390$ nm and $p_y=310$ nm, with $\alpha = 0.6$ and $h=120$ nm. \textbf{c}-\textbf{f}, Multipolar decomposition analysis: \textbf{c,e}, Calculated normalized scattering cross-sections and simulated reflectance (R) spectrum of a Sb$_2$S$_3$ metasurface with geometrical parameters of $p_{x,y}=310$ nm, $d_{x,y}=0.6 \, p_{x,y}$, and $h=120$ nm for the (\textbf{c}) amorphous and (\textbf{e}) crystalline phases. The constructive interference between the electric dipole (ED) and magnetic dipole (MD) modes at $\lambda_a = 560$ nm ($\lambda_c = 652$ nm) boosts the backward scattering intensity, and in turn, results in a reflectance peak in the case of A-Sb$_2$S$_3$ (C-Sb$_2$S$_3$). \textbf{d},\textbf{f}, Normalized magnetic field intensity with arrow surface of electric field (top panel), and normalized electric field intensity with arrow surface of magnetic field (bottom panel) for the metasurfaces in \textbf{b}(i) at (\textbf{d}) $\lambda_a = 560$ nm and (\textbf{f}) $\lambda_c = 652$ nm, respectively. }
\label{Fig_1}
\end{figure*}

To meet the challenges associated with plasmonic metasurfaces, recently, all-dielectric metasurfaces made of high-refractive-index materials supporting Mie-type resonances with electric dipole (ED) and magnetic dipole (MD) modes have been used for generating a full range of vivid and highly saturated structural colors desired for high-resolution display technologies \cite{ zhu2017resonant,yang2019ultrahighly, hemmatyar2019full, yang2020all}. However, these colors are fixed-by-design and cannot be tuned since the geometrical parameters of passive all-dielectric metasurfaces cannot be changed after fabrication. In order to enable active display applications, a real-time color tunability is essential.

To realize high-resolution structural color tunability in metasurfaces, several modulation techniques have been proposed. Some examples are liquid crystals in conjunction with plasmonic nanoantennae \cite{franklin2015polarization, olson2016high}, utilizing mechanically stretchable substrates integrated with plasmonic \cite{tseng2017two} and dielectric \cite{gutruf2016mechanically} nanoscatterers, changing the refractive index of the medium surrounded nanostructures \cite{king2015fano}, modifying the optical properties of the constituent magnesium-based nano-cavities of a hybrid plasmonic-dielectric platform via a chemical reaction \cite{chen2017dynamic}, and changing the polarization state of incident light \cite{yang2018polarization}.
Despite impressive advancements, these approaches can hardly meet the requirements for lightweight, flexible, durable, high-resolution, high-speed, and cost-effective dynamic color displays with high color contrast and saturation, multiple stable colors, and high refreshing rates. 

To overcome the existing shortcomings, chalcogenide phase-change materials (PCMs) \cite{wuttig2017phase,ding2019dynamic, abdollahramezani2020tunable, gholipour2013all,zhang2019broadband,taghinejad2021ito, rios2015integrated, abdollahramezani2021electrically, tian2019active, michel2019advanced,abdollahramezani2021dynamic,wu2021programmable, zheng2020nonvolatile,zhang2021electrically, wang2021electrical}, with optical properties (e.g., refractive index) that can be strongly modified upon applying an external stimulus (optical, electrical, or thermal), have been successfully used as tunable materials for color switching \cite{hosseini2014optoelectronic, tao2020phase, yoo2016multicolor, carrillo2019nonvolatile, de2020reconfigurable}. The advantages of PCM-based color-switching techniques over other counterparts originate from unique electrical and optical features of PCMs including nonvolatility, high index contrast, fast reversible switching speeds (10s-100s nanoseconds) between two stable phases, high durability (up to $10^{12}$ cycles), notable scalability (down to nm-scale sizes), good thermal stability (up to several hundred degrees), and adaptability with the CMOS fabrication technology \cite{abdollahramezani2020tunable}. Considering these unique features, single \cite{hosseini2014optoelectronic, tao2020phase} or multiple ultrathin films \cite{yoo2016multicolor} made of germanium antimony telluride (GST in short) and germanium telluride (GeTe) alloys in a multistack configuration with other dielectric and/or metallic films have been utilized for color switching \cite{carrillo2019nonvolatile}. In spite of the unparalleled properties of PCMs, these demonstrations suffer from the high absorption loss of GST and GeTe within the visible wavelength range, which results in low-quality-factor (low-Q) reflectance resonances. This, in turn, yields colors with low saturation, low color value (i.e., the reflectance value at the resonance peak) and purity in both amorphous and crystalline states of these PCMs.

To address these challenges, here we systematically design and experimentally demonstrate an actively tunable platform for color displays comprising all-dielectric metasurfaces formed by a geometrical arrangement of phase-change nanoellipsoids. We leverage a less explored class of PCMs, i.e., antimony trisulphide (Sb$_2$S$_3$) and antimony triselenide (Sb$_2$Se$_3$), exhibiting low-loss property in the visible spectral range \cite{ghosh1979optical, chen2015optical,dong2019wide,delaney2020new, delaney2021nonvolatile, liu2020rewritable}. Due to their high refractive indices, these materials support strong Mie-type ED and MD resonances. The sensitivity of these modes on refractive index enables high-resolution (up to $\sim$80,000 dots per inch (dpi)) phase-transition-based color switching with high saturation and purity \cite{hemmatyar2019full}. Moreover, owing to the polarization-sensitivity of the constituent asymmetric PCM nanopillars, we can encode two different colors into two mutually orthogonal polarization states of the incident light. This results in realization of a display with fixed geometrical parameters that can generate four different colors upon transition in the structural state of the contributed PCM. Finally, the integration of an electrically controlled transparent heater with the polarization-encoded phase-change meta-pixels, reported for the first time in this work, enables real-time reconfiguration for applications ranging from tunable full-color printing and displays, information encryption, and anticounterfeiting to wearable screens and electronic papers.

\section*{Results and Discussion}

Figure~\ref{Fig_1}a demonstrates the operation principle of a dynamic display formed by phase-change meta-pixels. Each meta-pixel is composed of a periodic array of rectangular unit cells, with different periodicity along x- and y-directions (i.e., $p_x$ and $p_y$ in Fig.~\ref{Fig_1}b (ii)), containing asymmetric elliptical Sb$_2$S$_3$ nanopillars on top of a glass substrate. The major and minor axes of the Sb$_2$S$_3$ nanopillars are proportional to the periodicity of the unit cell in the corresponding directions, i.e., $d_{x,y} = \alpha \, p_{x,y}$, in which $\alpha$ is fixed between 0 and 1. The height of the nanopillars ($h$) is constant for the fabrication preference. 
The reflected color upon normally incident x-polarized white light
can change by varying the geometrical parameters of the elliptical amorphous-Sb$_2$S$_3$ (A-Sb$_2$S$_3$) nanopillars or equivalently those of the unit cell (see the bottom-left display in Fig.~\ref{Fig_1}a). Upon phase transition, crystalline-Sb$_2$S$_3$ (C-Sb$_2$S$_3$) nanopillars with the same geometrical parameters and under the same illumination conditions generate colors that are different from those generated by their A-Sb$_2$S$_3$ counterparts (compare the top and bottom displays in Fig.~\ref{Fig_1}a). This phase-change color switching is attributed to the refractive index change of the constituent Sb$_2$S$_3$ meta-pixels upon transition between amorphous and crystalline phases.

To reveal the switching mechanism of, we performed the multipole decomposition analysis of the scattering spectrum of a Sb$_2$S$_3$ meta-pixel under white light illumination, as shown in Figs.~\ref{Fig_1}c-f (see Supporting Information Note~I and Fig.~S1 for more details). 
The analysis shows negligible contribution of the higher-order moments so that the optical response of the unit cell is governed by the electric dipole (ED) and magnetic dipole (MD) moments. In fact, the strong coupling between these refractive index-dependent ED and MD moments excited inside the Sb$_2$S$_3$ nanopillars with the directly reflected light yields the resonances in the reflectance spectra shown in Figs.~\ref{Fig_1}c,e. Therefore, the spectral position of these resonances, or equivalently the generated color by a meta-pixel under a specific polarized light, is determined by the refractive index of the Sb$_2$S$_3$ nanopillars. In addition to the color switching mechanism described above, the asymmetric nature of nanopillars can enable a polarization-based color switching in which one meta-pixel can generate different colors upon white light illumination with different polarization states (i.e., x- to y-polarization) in each phase of Sb$_2$S$_3$ (compare the left and right displays in Fig.~\ref{Fig_1}a). Therefore, one meta-pixel with fixed geometrical parameters can generate four different colors owing to the phase-change-tunability and polarization-sensitivity of the constituent Sb$_2$S$_3$ nanopillars.


\begin{figure*}[htbp]
\centering
\includegraphics[width=.85\linewidth, trim={0cm 0cm 0cm 0cm},clip]{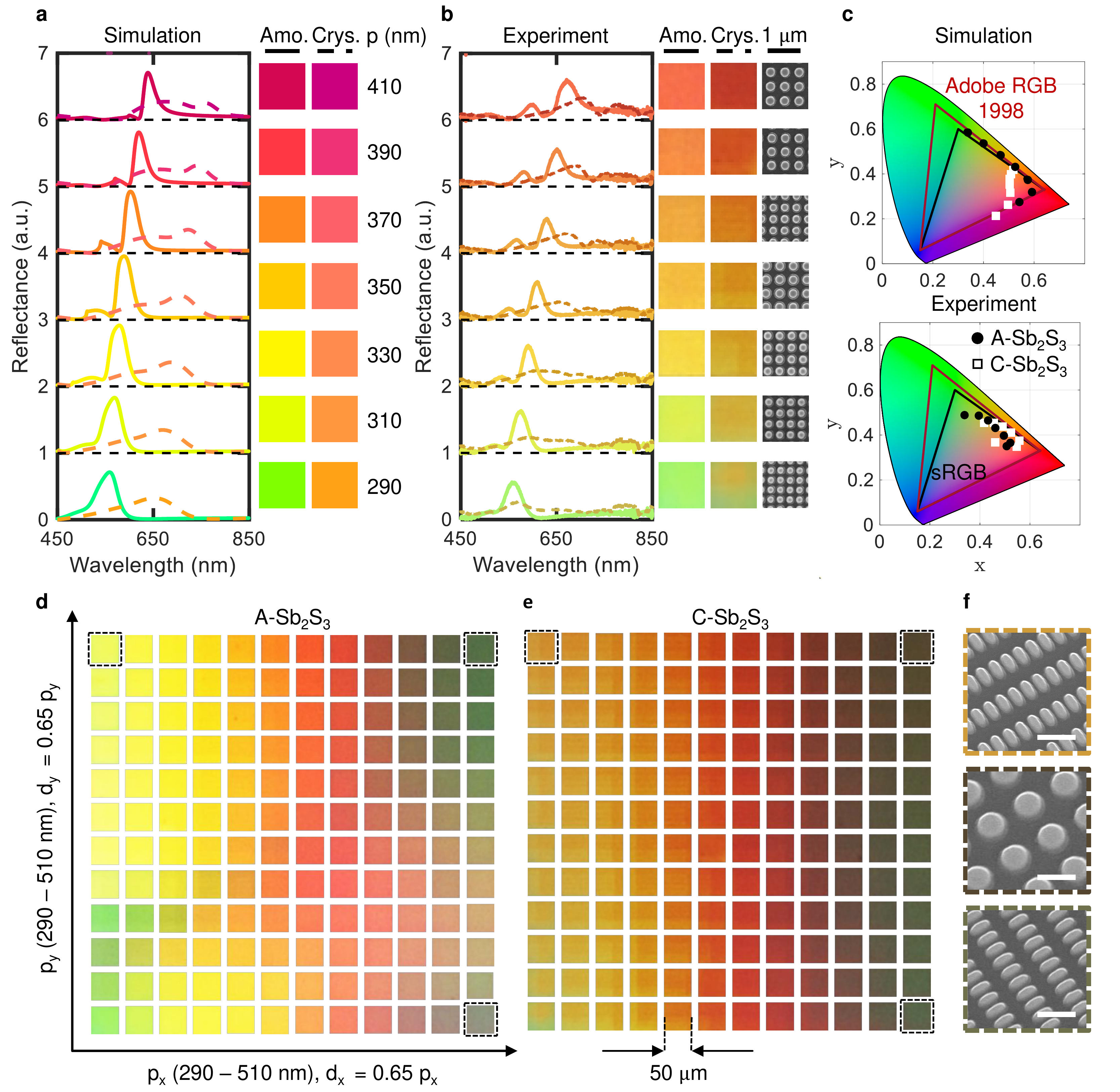}
\caption{\textbf{Simulation results and experimental characterization of polarization-encoded dynamic phase-change meta-pixels.} \textbf{a},\textbf{b}, Simulated (\textbf{a}) and experimental (\textbf{b}) reflectance spectra for the polarization-insensitive A-Sb$_2$S$_3$ (solid lines) and C-Sb$_2$S$_3$ (dashed lines) metasurfaces as well as their corresponding colors and SEM images for different periodicity ($p_x=p_y=p$). The curves are displaced vertically for better visibility and comparison. The diameter of Sb$_2$S$_3$ nanopillars varies as $d = 0.65\,p$. The sharp resonances observed in (\textbf{a}) and (\textbf{b}) are attributed to the interference between ED and MD modes inside the Sb$_2$S$_3$ nanopillars as shown in Figs.~\ref{Fig_1}(\textbf{c})-(\textbf{f}) causing the spectral position of these resonances become refractive-index-dependent. Therefore, red-shifting is observed upon phase transition of Sb$_2$S$_3$. \textbf{c}, Corresponding CIE 1931 chromaticity coordinates of the reflectance spectra shown in (\textbf{a},\textbf{b}) for A-Sb$_2$S$_3$ (black circles) and C-Sb$_2$S$_3$ (white squares). \textbf{d},\textbf{e}, The color palettes for the fabricated (\textbf{d}) A-Sb$_2$S$_3$ and (\textbf{e}) C-Sb$_2$S$_3$ meta-pixels considering different periodicity in x- and y-directions ($p_x$ and $p_y$, respectively) varying with 20 nm increments. \textbf{f}, SEM images showing magnified bird's eye views of three meta-pixels indicated by dashed boxes in (\textbf{d},\textbf{e}). The scale bar in (\textbf{f}) is 500 nm. The height of the Sb$_2$S$_3$ nanopillars is fixed at $h=120$ nm. The images of color pixels are captured through a $2.5\times$ objective lens with numerical aperture (NA) of 0.075.}
\label{Fig_2}
\end{figure*}

To study the effect of material phase on the generated color, we first consider polarization-insensitive meta-pixels with a square lattice ($p_x=p_y=p$) and circular ($d_x=d_y=d$) Sb$_2$S$_3$ nanopillars. We set the geometrical parameters as $h=120$ nm and $d = 0.65 \, p$ while we vary the periodicity of the unit cell from $p=290$ nm to $p=450$ nm, with a step of 20 nm, to cover a wide range of possible colors. The corresponding reflectance spectra obtained from full-wave simulations (see Method), and in turn, the generated colors are shown in Fig.~\ref{Fig_2}a. The method used for obtaining the associated colors with each reflectance spectrum is detailed in Supporting Information Note~II and Fig.~S3. As shown in Fig.~\ref{Fig_2}a, by increasing $p$, the spectral position of the resonance peaks for both amorphous (solid lines) and crystalline (dashed lines) states red-shift. Moreover, switching the phase of the material shifts the resonance peak to the longer wavelengths. This is due to positive refractive index contrast ($\Delta n_{\textrm{Sb}_{2}\textrm{S}_{3}} = n_{\textrm{C-Sb}_{2}\textrm{S}_{3}} - n_{\textrm{A-Sb}_{2}\textrm{S}_{3}} > 0$) within the visible wavelength range (see Supplementary Fig. S2c). The higher absorption loss of C-Sb$_2$S$_3$ mean that the nanopillars in the crystalline state (dashed lines in Fig. 2a) do not support equally strong and sharp resonances  as seen for A-Sb$_2$S$_3$ (solid lines).
 

To demonstrate the validity of our approach, we fabricated and characterized $50\times50$ $\mu$m$^2$ Sb$_2$S$_3$ meta-pixels with the same design parameters as those in Fig.~\ref{Fig_2}a (see Methods for fabrication and characterization details). The measured reflectance spectra, associated colors observed under the microscope as well as magnified top-view scanning electron micrographs (SEMs) of fabricated meta-pixels are shown in Fig.~\ref{Fig_2}b demonstrating an overall good agreement with the results obtained from simulations. To qualitatively analyze the performance of the presented color generation/switching mechanism in terms of saturation maintenance and hue variation, we display the generated colors in the amorphous (black circles) and crystalline (white squares) phases in the same International Commission on Illumination (CIE) 1931 chromaticity coordinates in Fig.~\ref{Fig_2}c. While for greenish and reddish colors, both simulation and experimental results demonstrate high saturation values (i.e., those markers close to the edge of the gamut), the purplish colors cannot be produced in the experiments. We attribute this to the presence of undesired secondary peaks observed in the reflectance spectra in Fig.~\ref{Fig_2}b for $p>390$ nm due to fabrication imperfections. A thorough quantitative study on the color gamut coverage, saturation, and hue for the case of Sb$_2$S$_3$ meta-pixels and an other low-loss PCM (i.e., Sb$_2$Se$_3$) is presented in Supplementary Note~III and Figs.~S4,5.

\begin{figure*}[htbp]
\centering
\includegraphics[width=.95\linewidth, trim={0cm 0cm 0cm 0cm},clip]{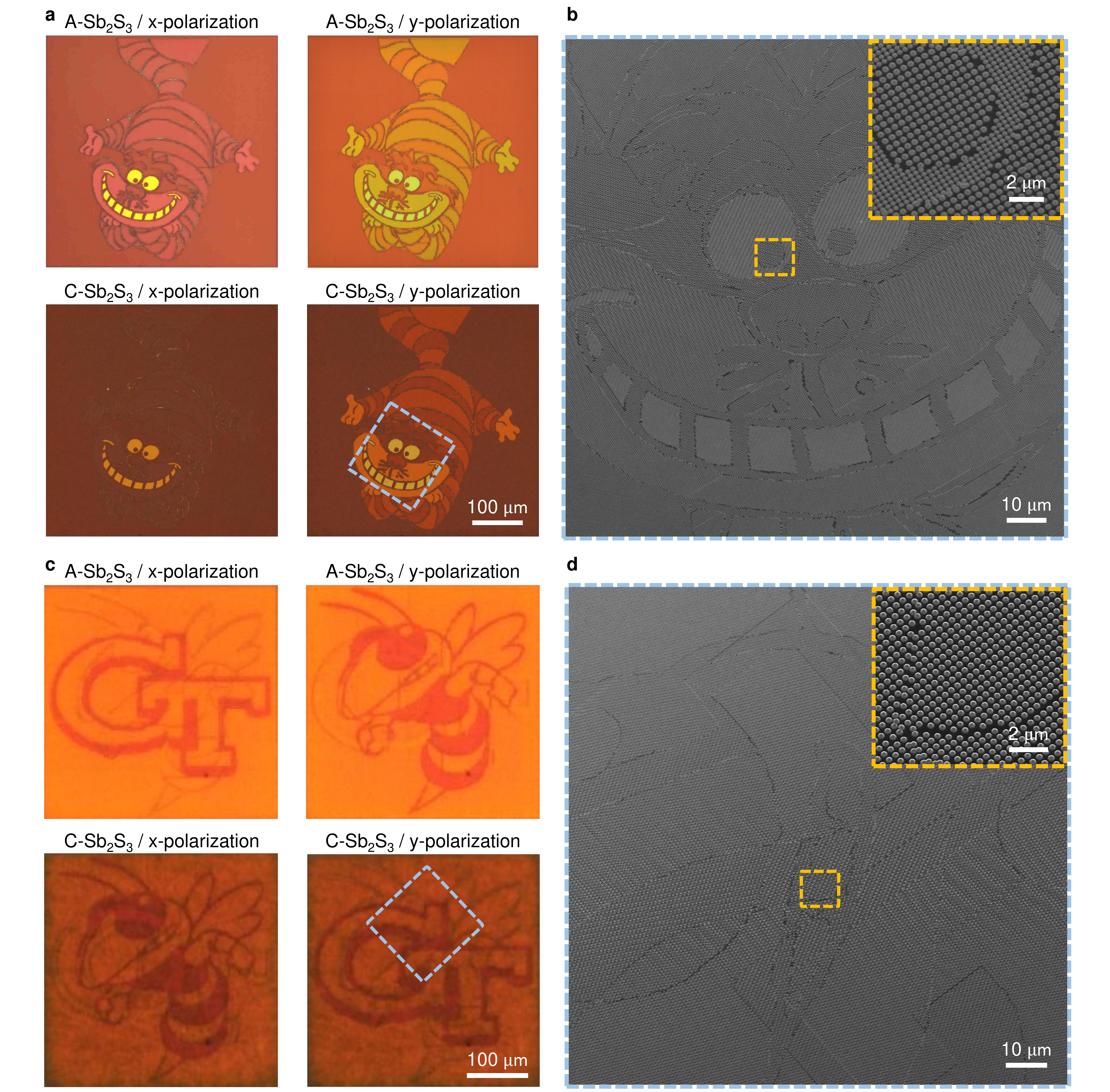}
\caption{\textbf{Dynamic displays enabled by polarization-sensitive Sb$_2$S$_3$ meta-pixels.} \textbf{a}, Reproduction of the image of The Cheshire Cat by A-Sb$_2$S$_3$ and C-Sb$_2$S$_3$ meta-pixels. For the case of A-Sb$_2$S$_3$, switching the polarization of incident white light changes the generated color throughout the image. This phenomenon is also observed for incident y-polarized light upon crystallization of A-Sb$_2$S$_3$. Under x-polarization, however, all parts of The Cheshire Cat body in the amorphous phase vanish except its teeth and eyes upon switching to the crystalline phase. This is also the case when changing the polarization of the incident white light from y- to the x-direction in the crystalline phase. \textbf{b}, The SEM image of the fabricated array of Sb$_2$S$_3$ nanopillars associated with the face of The Cheshire Cat indicated by the blue dashed box shown in (\textbf{a}). The magnified SEM image shown in the inset demonstrates that a meta-pixel containing only four Sb$_2$S$_3$ nanopillars is capable of generating the desired color justifying the high-resolution nature of the presented color-printing approach. \textbf{c}, Encryption of two images (i.e., Georgia Tech logo and symbol) into a display containing an engineered arrangement of Sb$_2$S$_3$ meta-pixels. One image can be switched to another either by changing the polarization of incident light in each phase, or by changing the phase of the Sb$_2$S$_3$ meta-pixels under the same polarization. The latter is the first experimental demonstration of encryption of two totally different images into the phase of the constituent material of meta-pixels. \textbf{d}, The SEM image of the fabricated Sb$_2$S$_3$ meta-pixels corresponding to the blue dashed box shown in (\textbf{c}). The design strategy and geometrical parameters of different parts of the images shown in (\textbf{a}-\textbf{d}) is explained in Supplementary Figs.~S15-17. The images in (\textbf{a}) and (\textbf{c}) are captured through $10\times$ (NA = 0.3) and $2.5\times$ (NA = 0.075) objective lenses, respectively.} 
\label{Fig_3}
\end{figure*}

\begin{figure*}[htbp]
\centering
\includegraphics[width=1\linewidth, trim={0cm 0cm 0cm 0cm},clip]{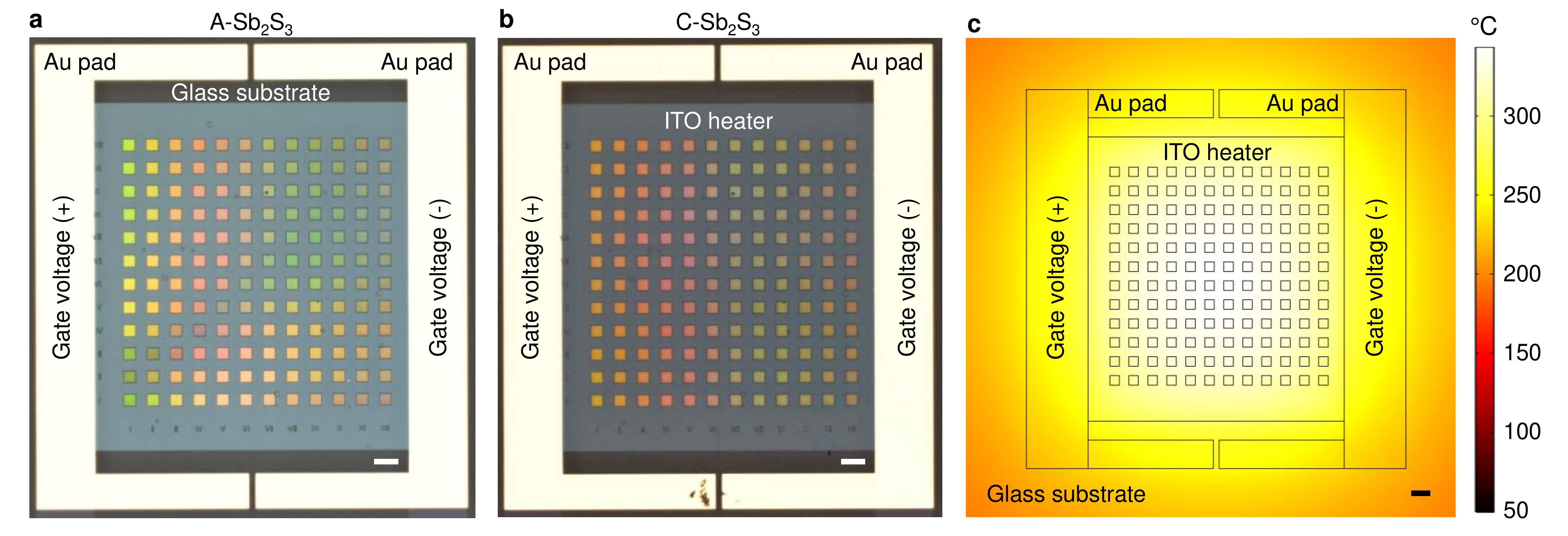}
\caption{\textbf{Electrically driven dynamic color display device integrating a transparent heater.} \textbf{a},\textbf{b}, The bright-field microscope images of the electrically tunable color palettes comprising $50\times50$ $\mu$m$^2$ (\textbf{a}) A-Sb$_2$S$_3$ and (\textbf{b}) C-Sb$_2$S$_3$ meta-pixels fabricated on a glass substrate and encapsulated by a 150 nm-thick film of SiO$_2$. The transparent heater is formed by fabrication of a 50 nm-thick ITO bridge connecting Au probing pads at the two ends on top of the SiO$_2$ film. The geometrical parameters of the Sb$_2$S$_3$ meta-pixels shown in (\textbf{a}) and (\textbf{b}) are similar to those used in Fig.~\ref{Fig_2} (\textbf{d},\textbf{e}). The images are captured using a $2.5\times$ objective (NA = 0.075). \textbf{c}, Simulated stationary temperature map in the cross section of Sb$_2$S$_3$ meta-pixels in the course of applying a 27 V electrical signal to the Au probing pads. The uniform heat distribution across the palettes ensures realization of large-scale displays with selective controllability of the material phase of all meta-pixels. The scale bars are 100 $\mu$m.}
\label{Fig_4}
\end{figure*}

In order to add polarization sensitivity to our color-switching approach, from now on, we also consider elliptical nanopillars in asymmetric unit cells with different periodicity in the x- and y-directions, i.e., $p_x$ and $p_y$, as shown in Fig.~\ref{Fig_1}b. By varying $p_x$ and $p_y$ from 290~nm to 510~nm with a 20-nm increment and a fixed ratio with respect to the major and minor axes of the nanopillars (i.e., $d_{x,y}=0.65 \, p_{x,y}$), we fabricate the color palettes shown in Figs.~\ref{Fig_2}d,e captured under x-polarized illumination. Measurement under y-polarization yields the same pattern flipped in $p_x$ and $p_y$ (results not shown here). The magnified bird’s eye view of three meta-pixels indicated by dashed boxes in Fig.~\ref{Fig_2}d,e are displayed in Fig.~\ref{Fig_2}f. The simulated palettes as well as a detailed analysis on the polarization-based and phase-change-based color switching approaches in the presented platform considering both Sb$_2$S$_3$ and Sb$_2$Se$_3$ meta-pixels are provided in Supplementary Note~IV and Figs.~S6-9. In addition to the ratio used in Fig.~\ref{Fig_2}d,e (i.e., $\alpha = 0.65$), we fabricate palettes of meta-pixels with other ratios of 0.45 and 0.55 and plot the captured images in Supplementary Fig.~S10. Moreover, based on the simulation results in Supplementary Fig.~S8, we design and fabricate palettes of Sb$_2$Se$_3$ meta-pixels with different ratios and display their microscopic images in Fig.~S11. The sensitivity of the generated colors to the polarization angle, incident angle, and different design parameters of the meta-pixels are analyzed in Supplementary Notes~V (Fig.~S12), VI (Fig.~S13), and VII (Fig.~S14), respectively.

The color switching enabled by the phase transition of Sb$_2$S$_3$ and polarization of the incident light can be employed for implementation of a dynamic display. According to the color palettes in Figs.~\ref{Fig_2}d,e, under x- (y-) polarization, a column (row) of different colors in the amorphous phase can be mapped onto a column (row) of relatively similar colors in the crystalline phase. We leverage this unique feature of the Sb$_2$S$_3$ meta-pixels for switching off some parts of an image while maintaining the colors of the remaining parts. As an illustrative example, the image of The Cheshire Cat is generated by A-Sb$_2$S$_3$ meta-pixels illuminated with x-polarized white light as shown in Fig.~\ref{Fig_3}a (i). Upon phase-transition to the C-Sb$_2$S$_3$, all parts of the body vanish, but the grinning and eyes remain (see Fig.~\ref{Fig_3}a (ii)). It is also the case when changing the polarization state from y to x in the crystalline phase. On the other hand, altering the polarization in amorphous phase as well as switching the material phase under y-polarization result only in a variation of colors in the components of the image. The SEM image of the fabricated array of Sb$_2$S$_3$ nanopillars associated with the face of The Cheshire Cat indicated by the blue dashed box shown in Fig.~\ref{Fig_3}a is displayed in Fig.~\ref{Fig_3}b. The magnified SEM image shown in the inset demonstrates that a meta-pixel containing only four Sb$_2$S$_3$ nanopillars can generate the desired color showing the high-resolution nature of the presented approach. The geometrical parameters of the meta-pixels used for the generation of Figs.~\ref{Fig_3}a,b are tabulated in Supplementary Fig.~S15.

Another interesting characteristic of our platform is its two degrees of freedom, i.e., phase-change and polarization-based color switching, at the same time. In fact, it is possible to darken (brighten) some parts of an image using the polarization-based control, while brightening (darkening) other parts using phase-change control of the meta-pixels. We benefit from this capability to encrypt two different images (i.e., Georgia Tech logo and symbol) into a display containing an array of Sb$_2$S$_3$ meta-pixels as shown in Fig.~\ref{Fig_3}c. One image can be switched to another one either by altering the polarization in each material phase, or by changing the phase of the Sb$_2$S$_3$ meta-pixels under a fixed polarization. While the former has been reported in previous works, the latter, to the best of our knowledge, is the first demonstration of encryption of two totally different images into the phase of the constituent materials of a meta-pixel. The SEM image of the fabricated Sb$_2$S$_3$ meta-pixels corresponding to the blue dashed box in Fig.~\ref{Fig_3}c is demonstrated in Fig.~\ref{Fig_3}d. The design strategy of the meta-pixels used for generating Figs.~\ref{Fig_3}c,d is provided in Supplementary Fig.~S16. Moreover, we demonstrate other examples of dynamic displays using Sb$_2$Se$_3$ meta-pixels in Supplementary Fig.~S17. For the case of Sb$_2$Se$_3$, we demonstrate the encoding and decoding of four different images into the A-Sb$_2$Se$_3$ (ON-state) and C-Sb$_2$Se$_3$ (OFF-state), respectively, under x- and y-polarizations. These capabilities can be used in many applications such as information coding, cryptography, high-density optical data storage, security encryption, and 3D displays.

In all experiments shown in Figs.~\ref{Fig_1}-\ref{Fig_3}, the phase-transition in our Sb$_2$S$_3$ meta-pixels is performed by using a bulky heater for a relatively long annealing time (see Methods for details). Though laser pulses can be used to expedite the conversion process \cite{liu2020rewritable}, the on-chip integration of high-power fast lasers is challenging if not impossible. This hinders the applicability of our approach for on-demand compact, high-resolution, fast, and on-chip display. To promote the presented approach to a practical paradigm, we must electrically convert the Sb$_2$S$_3$ meta-pixels. Recently, electrical switching of PCMs based on Joule heating has been successfully demonstrated using metal micro-heaters \cite{abdollahramezani2021electrically,zhang2021electrically, wang2021electrical}. However, none of these platforms is suitable for structural color generation due to the excessive loss of their constituent material in the visible range. Thanks to their reduced optical loss, micro-heater formed in transparent conductive oxides hold the promise to enable next-generation dynamic structural colors. 

As a proof-of-concept demonstration, we leverage an indium tin oxide (ITO) heater to electrically reconfigurequality if the the phase-change meta-pixels without compromising the quality of the generated colors. To this end, the fabricated palettes in Figs.~\ref{Fig_2}d,e are first encapsulated by a SiO$_2$ layer followed by fabrication of a 50 nm-thick indium tin oxide (ITO) bridge connecting two gold (Au) probing pads at the two ends on top of the SiO$_2$ film (see Figs.~\ref{Fig_4}a,b and Methods for fabrication details). The electro-thermal simulation in Fig.~\ref{Fig_4}c illustrates that a fairly uniform heat distribution can be realized across the whole area of the display upon applying the voltage pulse ensuring simultaneous and uniform conversion of all palettes. Such a Joule heating platform offers the precise electrical control of the intermediate phases of PCMs (beyond amorphous and crystalline) which is critical for realization of multicolor displays, a key attribute of our approach. We further investigate the potential of ITO-based micro-heater for reversible switching of colors in Supplementary Information Figs.~S19-22.

\section*{Conclusion}

In summary, we demonstrated here a new platform for generating and switching high-efficiency, high-saturation, and wide-gamut structural colors using switchable meta-pixels by employing PCM-based metasurfaces made of low-loss and less explored Sb$_2$S$_3$ and Sb$_2$Se$_3$ nanopillars. Upon the nonvolatile phase-transition of the constituent PCM, the generated color in the amorphous phase switches to a distinctive stable color in the crystalline phase. In addition, the properly designed asymmetric characteristics of elliptical nanopillars enable polarization-based color switching. Combining these two tuning mechanisms, we systematically designed a single-layer meta-pixel capable of producing four different colors. This can be extended to the realization of multi-color artificial images by gradually changing the crystallinity of the constituent PCMs and/or the incident polarization angle. We also showed that by engineering the arrangement of PCM-based nanopillars, features like image switching, ON/OFF switching, and color shading can be realized. More interestingly, we experimentally demonstrate, for the first time, an electrically driven micro-scale display by integration of an optically-transparent heater to our color without compromising the color quality. We believe that this research provides a significant step towards the realization and commercialization of compact metaphotonic devices for applications like full-color dynamic displays, information storage, image encryption, and anti-counterfeiting. 

\section*{Acknowledgements}

The work was primarily funded by the Office of Naval Research (ONR) (N00014-18-1-2055, Dr. B. Bennett) and by the Air Force Office of Scientific Research MURI program. The support of the UK's Engineering and Physical Science Research Centre is gratefully acknowledged, through ChAMP–Chalcogenide Advanced Manufacturing Partnership (EP/M015130/1). The Stanford authors acknowledge partial support from the Stanford Graduate Fellowship, from the Nonvolatile Memory Technology Research Initiative (NMTRI), and from Draper Labs. This work was performed in part at the Georgia Tech Institute for Electronics and Nanotechnology (IEN), a member of the National Nanotechnology Coordinated Infrastructure (NNCI), which is supported by NSF (ECCS1542174).

\section*{Disclosures} The authors declare no conflicts of interest.

\section*{Methods}

\noindent \textbf{Sample fabrication.}
The fabrication flow for the Sb$_2$S$_3$ metasurface and integrated transparent heater of the meta-display is illustrated in Fig.~S23. A Sb$_2$S$_3$ film of nominally 130 nm thickness is first sputtered on a cleaned fused silica substrate from a stoichiometric target followed by the deposition of a 15-nm thick ZnS:SiO$_2$ film serving as a protective layer to prevent oxidation and elemental loss of Sb$_2$S$_3$ undergoing the heating process. Next, the sample is coated with a layer of hydrogen silsesquioxane (HSQ) negative e-beam resist and a thin water-soluble conductive layer of ESpacer to hamper the charge accumulation during the writing process. E-beam lithography is them performed to define the nanopillar pattern in each 50$\times$50 $\mu$m$^2$ meta-pixel. After washing out ESpacer using DI water, the exposed photoresist is developed by subsequently immersing it in a bath of 25\% tetramethylammonium hydroxide (TMAH) and rinsing with gently flowing DI water. Inductively couple plasma reactive ion etching (ICP-RIE) is performed with a gas mixture of Ar:CF$_4$ with the etching rate of $\sim$ 75 nm/min to form nanostructure patterns. The etching process is conducted through two 1-min cycles with a long-enough cooldown break in between. Right after the etching, a 15 nm protective layer of SiO$_2$ is grown on the sample using atomic layer depostion (ALD) at 100 $^{\circ}$C, which is low enough to prevent the crystallization of Sb$_2$S$_3$. To convert the material state of Sb$_2$S$_3$ to the crystalline phase, the sample is annealed at 270 $^{\circ}$C for 10 mins in a chamber filled with an ultrahigh pure Ar gas.

To realize the electrically-driven display, the fabricated sample (excluded from the annealing process) is first transferred to the ALD system to deposit a 200-nm thick layer of thermal SiO$_2$ as a supporting substrate for the integrated heater. After defining the pattern of the ITO bridge in the polymethyl methacrylate (PMMA)-coated sample using e-beam lithography, a 50 nm-thick layer of ITO is deposited by the RF-magnetron sputtering from an indium oxide/tin oxide (In$_2$O$_3$/SnO$_2$ with 90/10 wt \%) target in an argon/oxygen plasma atmosphere. The prolonged nature of the deposition facilitates the crystallization of ITO necessary for the formation of a uniform conductive layer enabling spatially consistent heat generation. After the lift-off process, to further enhance the electrical conductivity of the ITO film, post-deposition annealing under a mild flow of oxygen, which also reduces the optical loss of ITO, is conducted at 200 $^{\circ}$C for 30 mins. This temperature is low enough to preclude crystallization of as-deposited Sb$_2$S$_3$, . Two Au/Ti (250/20 nm) electrodes are formed at the two ends of the ITO bridge through subsequent e-beam lithography and e-beam evaporation processes. After the lift-off process, in the final step, a 100 nm layer of SiO$_2$ is grown to prevent the failure of the heater caused by the electric breakdown of the air at the sharp corners of the device. To fully transform the Sb$_2$S$_3$ phase from amorphous to crystalline based on the Joule heating process, a 32 V long-enough (1 min) pulse is applied to the integrated heater using a source measurement unit (Keithley 2614B).

\noindent \textbf{Optical measurements.} To investigate the optical response of the fabricated meta-displays, bright-field optical imaging and reflection spectra measurements of the color palettes are conducted. Optical images are captured using a conventional upright bright-field reflection microscope (Nikon ECLIPSE L200, Nikon Inc.) equipped with a high-definition color camera head (DS-Fi2) and a 50 W halogen lamp light source. To observe different colors of The Cheshire Cat and Georgia Tech logo and symbol images under different polarization states of incident white light, the corresponding images are magnified with a 10$\times$ objective lens (NA = 0.3) and a 2.5$\times$ objective lens (NA = 0.075), respectively, under illumination of polarized light in both orthogonal directions.
The optical spectra ($\lambda$ = 450-850 nm) are measured in reflection mode using a home-built microscope set-up equipped with a 75 W broadband xenon source (Newport) and a UV-visible-near infrared (NIR) spectrometer (USB 2000+, Ocean Optics Inc.). The polarized light illuminates a colour palette at normal incidence through an achromatic 10$\times$ objective lens (NA = 0.25) and is collected through the same objective and back into the spectrometer and a CCD camera. The measured reflectance spectra are normalized to the reflected light from an aluminium-coated mirror. All measurements are carried out at room temperature ($\sim$ 25 $^{\circ}$C).

\noindent \textbf{Numerical simulations.}
The full-wave simulations of the reflectance spectra of the metasurfaces are performed using the commercial software Lumerical Solutions based on the finite-different time-domain (FDTD) technique. The periodic boundary condition is used in the x- and y-directions to mimic the periodicity, while perfectly matched layers are used in the z-direction (top and bottom layers) to model the free space. The refractive index of the glass substrate is set at 1.46 for the entire wavelength range. The dispersive optical constants of PCMs obtained from spectroscopic ellipsometry measurements shown in Supplementary Fig. S2 are incorporated into simulations.

\noindent \textbf{Electro-thermal simulations.} 
A three-dimensional finite element method (FEM) simulation is performed in the software package COMSOL Multiphysics to simulate the Joule heating and heat dissipation effects in the electrified hybrid display. In our simulations, we consider certain assumptions and boundary conditions to mimic the experimental conditions. The multiphysics problem is solved through coupling of an electric currents (ec) module to a heat transfer in solid (ht) physics model. Material properties used for fused silica, Ti, Au, and ITO are adopted from the available references \cite{lide2004crc,rios2018controlled}. The electrical conductivity of ITO obtained from the four-point probe measurement is set at 1.42$\times$10$^4$ S/m. The thermal conductivity, density, and heat capacity of Sb$_2$S$_3$ are 1.16 W/m·K, 4600 kg/m$^3$, and 120 J/mol.K, respectively \cite{liu2020rewritable}. The ec module is applied to the ITO bridge and electrodes. Electric insulation are assigned to all boundaries except for the two endfaces of the bridge where normal current density and electric ground are applied. The ht physics model is assigned to all domains. The convective cooling boundary condition with an ambient temperature of 20 $^{\circ}$C and the heat transfer coefficient of 5 W/m$^2$.K is used at the top and bottom surfaces. Open boundary condition is applied to the walls of the substrate in the lateral directions.

\bibliography{apssamp}


\clearpage
\newpage


\setcounter{figure}{0}
\makeatletter
\renewcommand{\fnum@figure}{\figurename~S\thefigure}
\makeatother
\setcounter{equation}{0}
\renewcommand{\theequation}{S.\arabic{equation}}



\newpage

\section*{Supplementary Information}

\section{Multipolar decomposition}

Electromagnetic properties of the nanoparticles in the arrays are numerically studied by using the commercial software CST Microwave Studio$^{\rm TM}$. In the canonical basis we perform a multipole expansion of the scattered field of the hybrid nanoparticles into vector spherical harmonics, which form a complete and orthogonal basis allowing the unique expansion of any vectorial field. To calculate electric (a$_E$(l,m)) and magnetic (a$_M$(l,m)) spherical multipole coefficients, we project the scattered electric field $\mathbf{E}_{sca}$ on a spherical surface, enclosing the nanoparticles centered at the symmetric point of the nanodisc, onto vector spherical harmonics based on the following relations \cite{jackson1999classical, grahn2012electromagnetic}:
\begin{equation}\label{multipole}
\begin{split}
a_E(l,m)=&\frac{(-i)^{l+1}kR}{h_l^{(1)}(kR)E_{0}\sqrt{\pi (2l+1)(l+1)l}}\\
&\int_{0}^{2\pi}\int_{0}^{\pi}Y^{*}_{lm}(\theta, \phi)\mathbf{r}\mathbf{E}_{sca}(\mathbf{r})\sin\theta d\theta d\phi,
\end{split}
\end{equation}
\begin{equation}
\begin{split}
a_M(l,m)=&\frac{(-i)^{l}kR}{h_l^{(1)}(kR)E_{0}\sqrt{\pi (2l+1)}}\\
&\int_{0}^{2\pi}\int_{0}^{\pi}\mathbf{X}^{*}_{lm}(\theta, \phi)\mathbf{E}_{sca}(\mathbf{r})\sin\theta d\theta d\phi,
\end{split}
\end{equation}
where $R$ is the radius of the enclosing sphere, $k$ is the wavenumber, $h_l^{(1)}$ is the Hankel function with the asymptotic of the outgoing spherical wave, $E_{0}$ is the amplitude of the incident wave, $Y^{*}_{lm}$ and $\mathbf{X}^{*}_{lm}$ are scalar and vector spherical harmonics. The integers $l$ and $m$ describe the order of the multipole (dipole, quadrupole, ...) and the amount of the z-component of angular momentum that is carried per photon, respectively. Due to the azimuthal symmetry of the nanoparticles under normal excitation, the amplitude of the scattering coefficients with opposite $m$ indices are identical, i.e., $a_{E,M}(l,m)=a_{E,M}(l,-m)$.

\section{Color generation}

To achieve generated colors, the International Commission on Illumination (CIE) XYZ tristimulus values corresponding to the reflection spectra are calculated as \cite{hemmatyar2019full}:
\begin{gather}
X = \frac{1}{k} \int{I(\lambda) R(\lambda) \bar{x}(\lambda) d\lambda},\nonumber\\
Y = \frac{1}{k} \int{I(\lambda) R(\lambda) \bar{y}(\lambda) d\lambda}, \\
Z = \frac{1}{k} \int{I(\lambda) R(\lambda) \bar{z}(\lambda) d\lambda}\nonumber
\end{gather}
where $k$ is the normalization factor, \(I(\lambda)\) is energy distribution of the reference light; \(R(\lambda)\) is the reflection spectrum obtained from the designed mestasurface under illumination; and \(\bar{x}(\lambda)\), \(\bar{y}(\lambda)\), and \(\bar{z}(\lambda\)) are the CIE 1931 standard color-matching functions (see Figure~S2a). These chromaticity functions are then normalized as $x=X/(X+Y+Z)$ and $y=Y/(X+Y+Z)$, which fall between 0 to 1, to represent the colors in the CIE 1931 chromaticity diagram.

\section{Quantitative analysis on color gamut coverage, saturation maintenance and hue variation}

As shown in Fig.~S4d-f, by increasing $p$, the spectral position of the reflectance resonances for both amorphous (solid lines) and crystalline (dashed lines) states red-shifts. The spectral position of each of these resonances is dependent on the refractive index of the constituent phase-change material (PCM) owing to the interference between ED and MD modes inside the PCM nanopillars as will be discussed later. Therefore, by switching the state of the nanopillars from amorphous to crystalline, the central wavelengths of the resonances red-shift in the cases of Sb$_2$S$_3$ and Sb$_2$Se$_3$ (see Fig.~S4d,e), and blue-shift in the case of GeSe$_3$ (see Fig.~S4f) because $\Delta n_{\textrm{Sb}_{2}\textrm{S}_{3}}$, $\Delta n_{\textrm{Sb}_{2}\textrm{Se}_{3}} > 0$, while $\Delta n_{\textrm{GeSe}_{3}} < 0$ (with $\Delta n = n_{\textrm{C-PCM}} - n_{\textrm{A-PCM}}$) within the visible wavelength range (see the refractive indices in Fig.~S2a,b). The actual shift of the resonance wavelengths are $|\Delta\lambda_{\textrm{Sb}_{2}\textrm{S}_{3}}| < 180$ nm, $|\Delta\lambda_{\textrm{Sb}_{2}\textrm{Se}_{3}}| < 200$ nm, and $|\Delta\lambda_{\textrm{GeSe}_{3}}| < 70$ nm (see Fig.~S4d-f). The relative strength of the wavelength shifts in these PCMs, i.e. $|\Delta\lambda_{\textrm{Sb}_{2}\textrm{Se}_{3}}| > |\Delta\lambda_{\textrm{Sb}_{2}\textrm{S}_{3}}| > |\Delta\lambda_{\textrm{GeSe}_{3}}|$, is attributed to the relative strength of the change in the real part of their refractive indices upon the phase transition between amorphous and crystalline, i.e. $|\Delta n_{\textrm{Sb}_{2}\textrm{Se}_{3}}| > |\Delta n_{\textrm{Sb}_{2}\textrm{S}_{3}}| > |\Delta n_{\textrm{GeSe}_{3}}|$, as shown in Supplementary Fig.~S1. On the other hand, the sharpness of the reflectance resonances in Fig~S4d-f is mainly dependent on the PCM extinction coefficient shown as the dashed curves in Fig.~S4e,f. In the case of Sb$_2$S$_3$ nanopillars, the high-efficiency resonances (i.e., those with high reflectance value at the resonance peak) in the low-loss amorphous phase are damped upon the transition to the crystalline phase with higher absorption loss (compare solid and dashed curves in Fig.~S4d). This high absorption loss arises for both amorphous and crystalline Sb$_2$Se$_3$ nanopillars, resulting in relatively low-efficiency reflectance resonances (see Fig.~S4f). In contrast, GeSe$_3$ nanopillars remain very low-loss across the entire visible range for both the amorphous and crystalline phases, yielding high-efficiency resonances in both cases (see Fig.~S4f).

For a quantitative comparison between the presented three PCMs in terms of color generation/switching, Figs.~\ref{Fig_2}g-i show the generated colors in the amorphous (black circles) and crystalline (white squares) phases in the same International Commission on Illumination (CIE) 1931 chromaticity coordinates for the three PCMs in top panels, and their corresponding hue and saturation values for amorphous (solid-circle lines) and crystalline (dashed-square curves) phases in the bottom panels. The approach of calculating the CIE XYZ tristimulus of the reflectance spectra and their corresponding hue and saturation values are given in the Supporting Information Note~I. In terms of the color gamut coverage, the calculated color gamut area for A-Sb$_2$Se$_3$ (C-Sb$_2$Se$_3$) is around 98.3\% (43.4\%) of the standard RGB (sRGB) and 72.9\% (32.2\%) of the Adobe RGB, from Fig.~\ref{Fig_2}g. The color gamut area for the case of A-Sb$_2$Se$_3$ (C-Sb$_2$Se$_3$) is around 70.1\% (33.3\%) of the sRGB, and 52\% (24.7\%) of the Adobe RGB (from Figure~\ref{Fig_2}h). For the case of A-GeSe$_3$ (C-GeSe$_3$), a full-range of colors with gamut area of 57.8\% (90.8\%) of the sRGB, and 42.9\% (67.3\%) of the Adobe RGB can be obtained (from Figure~\ref{Fig_2}i). Therefore, in terms of color gamut area, Sb$_2$S$_3$ and GeSe$_3$ have almost the same performance, yet better than Sb$_2$Se$_3$. Moreover, these results show that our all-dielectric PCM-based metasurfaces can generate a wide color gamut larger than the state-of-the-art plasmonic colors ($\sim45\%$ of sRGB \cite{rezaei2019wide}) for A-Sb$_2$S$_3$, A-Sb$_2$Se$_3$ and A/C-GeSe$_3$ cases.

In the RGB color-mixing model, the hue (H) is defined as the proportion of the dominant wavelength (resonance wavelength in this case) with respect to other wavelengths in the reflected light and is independent of the intensity of the light. It simply indicates the "perceived color" by the human eyes and ranges from $0^\circ$ to $360^\circ$, in which $0^\circ$ (and $360^\circ$), $120^\circ$ and $240^\circ$ represent pure red, pure green, and pure blue, respectively (See Supplementary Fig.~S3 for more details). The saturation, on the other hand, is defined as the ratio of the intensity of the reflected light at the resonance wavelength (associated to the perceived color) to that of the incident white light, simply indicating the purity of a color and ranging from 0\% to 100\%. Considering this definition, the narrower the bandwidth of the reflectance resonance, the higher the saturation of the generated color. In the content of color switching between two phases, the performance measure is achieving two highly-saturated colors in both phases with a maximum hue variation ($\Delta \textrm{H} = \textrm{H}_{\textrm{C-PCM}} - \textrm{H}_{\textrm{A-PCM}}$) upon switching. To analyze the performance of the presented phase-transition-based color-switching approach, the hue and saturation values of the simulated colors in Fig.~S4d-f are plotted in the bottom panels in Fig.~S4g-i for both amorphous (solid-circle lines) and crystalline (dashed-square curves) phases of the PCMs. In terms of saturation preservation upon phase transition, GeSe$_3$ shows high-saturation values for both amorphous and crystalline cases (due to sharp reflectance resonances), while Sb$_2$S$_3$ shows highly saturated colors only in the amorphous phase. Sb$_2$Se$_3$, however, demonstrates a median level of saturation values in both amorphous and crystalline phases due to the wide reflectance resonances. With regards to hue variation, the hues of the generated colors in Sb$_2$S$_3$ and GeSe$_3$ cases change by varying $p$ in both amorphous and crystalline states while maintaining $\Delta \textrm{H} < 80^\circ$ upon phase transition. One may use this feature to switch the coloration of pixels of an image individually with each pixel being a Sb$_2$S$_3$ or GeSe$_3$ metasurface formed by of an array of down to 5$\times$5 or 6$\times$6 nanopillars \cite{hemmatyar2019full}. In the case of Sb$_2$Se$_3$, however, by changing $p$, all the varying hue values in the amorphous phase switch to an almost fixed hue in the crystalline phase. Using this property, one can turn off all the pixels of an image on a display comprising Sb$_2$Se$_3$ metasurfaces (i.e., pixels) by switching the phase of the Sb$_2$Se$_3$ nanopillars from the amorphous state (ON-state) to the crystalline state (OFF-state). This is a unique feature that is absent in other approaches in previous works, e.g., the polarization-sensitive color-switching approach \cite{yang2018polarization}.

To provide a comparison between Sb$_2$S$_3$, Sb$_2$Se$_3$, and GeSe$_3$ metasurfaces for color switching applications, a spider chart is shown in Fig.~S5. The figure of merit (FOM) is defined as the maximum variation of the hue over the refractive index change open phase transition between amorphous and crystalline, i.e. $|\Delta \textrm{Hue}|/|\Delta n|$ in which $\Delta n = n_\textrm{A}(\lambda_\textrm{A}) - n_\textrm{C}(\lambda_\textrm{C})$), with $n_\textrm{A}(\lambda_\textrm{A})$ and $n_\textrm{C}(\lambda_\textrm{C})$ being the index of refraction in the amorphous and crystalline phases and at the corresponding resonance wavelengths ($\lambda_\textrm{A}, \lambda_\textrm{C}$), respectively. While high FOM is desirable, the saturation and value (i.e. the reflectance value at the resonance peak) of the colors in both amorphous and crystalline phases should be as high as possible. Considering all these performance measures, GeSe$_3$ demonstrates superior properties over Sb$_2$S$_3$ and Sb$_2$Se$_3$ when switching from a color associated with a reflectance spectrum with a resonance peak at $\lambda = 600$~nm (chosen as the middle wavelength in the visible range from 400~nm to 800~nm) in the amorphous phase, to another color in the crystalline phase.

\section{Polarization-sensitive dynamic color generation}

To add the polarization-sensitivity to our color-switching approach, we also consider elliptical nanopillars in asymmetric unit cells with different periodicities in the x- and y-directions, i.e. $p_x$ and $p_y$, Fig~1b. By varying $p_x$ and $p_y$ with a fixed ratio with respect to the major and minor axes of the nanopillars (i.e., $d_{x,y}=\alpha \, p_{x,y}$), we generate the color palettes shown in Fig.~S6a,d,g, for the case of Sb$_2$S$_3$ ($p_{x,y}$ range from 310~nm to 470~nm with 40-nm increments), Sb$_2$Se$_3$ ($p_{x,y}$ range from 200~nm to 400~nm with 50-nm increments), and GeSe$_3$ ($p_{x,y}$ range from 270~nm to 430~nm with 40-nm increments), respectively (see Supplementary Fig.~S4-S6 in the for full color palettes). In each figure, the top (bottom) panels show the colors generated by the x-polarized (y-polarized) incident white light for amorphous (left panels) and crystalline (right panels) cases. While Sb$_2$S$_3$ and GeSe$_3$ metasurfaces can generate a full palette considering both amorphous and crystalline phases (see Fig.~S6a,g), respectively, Sb$_2$Se$_3$ metasurfaces cannot generate bluish colors (see Fig.~S6d). This stems from the high optical loss of Sb$_2$Se$_3$ within the blue range of the visible wavelengths (see Fig.~S2a,b). It is also clear that the y-polarization palettes can be obtained by transposing the x-polarization palette, i.e., replacing each (j,i) element with corresponding (i,j) element. However, this is not the case for amorphous and crystalline palettes in Fig.~S6a,d,g since the crystalline palettes contain completely different colors from those in the amorphous palettes. This shows the advantage of using PCMs as the number of colors in the phase-transition-based color-switching approach is twice as many as those in the polarization-based approach.

To analyze the effect of polarization-sensitivity in both amorphous and crystalline cases on the reflected colors, we select five metasurfaces for each PCM with geometrical parameters in the dashed boxes in Fig.~S6a,d,g, and plot the corresponding simulated reflectance spectra with their hue and saturation values in the inset in Fig.~S6b,e,h, respectively. It is seen that by increasing $p_y$ in each box, the central wavelength of the reflectance resonances does not experience a considerable shift for the x-polarization (see the top panels in Fig.~S6b,e,h). This leads to almost unchanged hue values for the corresponding colors, which in turn results in a limited trajectory in the corresponding color gamuts shown in the top panels of Fig.~S6c,f,i in which black circles (white squares) represent the colors in amorphous (crystalline) phase. In contrast, it is observed that increasing $p_y$ results in a tangible redshift in the reflectance spectra for the y-polarization for all PCMs (see the bottom panels in Fig.~S6b,e,h). This redshift results in a relatively large hue change in all cases, except C-Sb$_2$Se$_3$, as the corresponding color gamuts in the bottom panels of Fig.~S6c,f,i demonstrate. The simulated full color palettes as well as their corresponding gamuts are provided in Figs.~S7-9. Based on these simulation results, we designed and fabricated palettes of Sb$_2$S$_3$ and Sb$_2$Se$_3$ meta-pixels with different ratios and display their corresponding microscopic images in Fig.~S12 and 13, respectively.

Finally, in the Supplementary Note~V, we show that by continuously varying the incident polarization angle ($\varphi$) one can enable dynamic color tuning (See Figure~S12).

\section{Sensitivity to the incident polarization angle}

To analyze the effect of the variation of the incident polarization angle ($\varphi$) on the reflected colors, we select one metasurface for each type of PCMs with geometrical parameters shown in Fig.~S12a,d,g and change $\varphi$ from $0^\circ$ (y-polarization) to $90^\circ$ (x-polarization). The reflectance spectra of these metasurfaces for $\varphi=0^\circ$ and $\varphi=90^\circ$ for both amorphous and crystalline states are plotted in Fig.~S12c,f,i. In both amorphous and crystalline states, a resonance shift of at least 100 nm is observed, which enables us to dynamically tune the reflected colors by varying the incident polarization angle. This polarization-based color tunability is demonstrated in the colors in Fig.~S12b,e,h, which are generated through varying $\varphi$ from $0^\circ$ to $90^\circ$ in a step of $15^\circ$. The colors in Fig.~S12b,e,h and their corresponding CIE diagrams show that using Sb$_2$S$_3$ and GeSe$_3$ metasurfaces, one can tune the colors from green to reddish purple to blue, while Sb$_2$Se$_3$ can enable color tuning from dark green to red to purple.

\section{Sensitivity to the incident angle}
To analyze the effect of the incident angle ($\theta$) on the reflectance spectrum of a metasurface (Fig.~1b), we select a metasurface with Sb$_2$S$_3$ nanopillars, as shown in Fig.~S13a,b, and vary the angle of the incident light from $\theta = 0^{\circ}$ to $\theta = 30^{\circ}$. Fig.~S13c,d show the reflection spectra for amorphous Sb$_2$S$_3$, and Fig.~S13e,f show the results of crystalline Sb$_2$S$_3$, with TE- and TM-polarized light, respectively. In the case of TE-polarized light incident on amorphous Sb$_2$S$_3$ (Fig.~S13c), the incident angle has a small impact on the reflection spectrum. The intensity of the reflected is reduced by 20\% when $\theta$ approaches $5^{\circ}$, but the reflection spectra does not suffer any redshift. The spectra resulted from the crystalline Sb$_2$S$_3$ experiences a redshift of more than 100 nm and is less intense and is less intense compared to the amorphous case, but these spectra remain largely unaffected by the incident angle variation.

In the case of TM-polarized light on amorphous Sb$_2$S$_3$, a much greater dependence on $\theta$ is observed from Fig.~S13d,f. As $\theta$ increases, two effects can be seen from these figures: 1) the initial peak at $\theta=0 ^{\circ}$ seen begins to lose intensity and experiences a redshift, and 2) a new peak forms and becomes more pronounced, both as $\theta$ goes beyond $5^{\circ}$. When Sb$_2$S$_3$ is crystalline in this case, no considerable changes are observed for $0^{\circ}<\theta<20^{\circ}$ after which the peak redshifts by about 100 nm at $\theta = 30 ^{\circ}$. In addition, for $\theta > 20^{\circ}$, the second peak that was observable in the amorphous case is not seen in the crystalline case.  These results are not surprising; the reflectance of these metasurfaces is largely due to ED and MD resonances that are supported by the nanopillar structures, and the ED resonances are the dominating resonances seen in the reflectance spectra. Since the component of the electric field parallel to the top surface of the Sb$_2$S$_3$ nanopillars does not change in the case of the obliquely incident TE-polarized light (See Fig.~S9a), the incident angle should not have a major effect on the output spectra. Likewise, since the this component of the electric field changes in the case of obliquely incident TM-polarized light (See Fig.~S13b), we should see a greater impact of varying $\theta$ on the resulting spectra.

\section{Influence of different design parameters}
Analysis must also be done to determine the effects that the physical dimensions of the nanopillars have on the reflection spectrum. Fig.~S14a,b,c show the reflectivity spectrum of a Sb$_2$S$_3$ array with nanopillars Fig~1b of varying heights (h), periods (p), and diameters (d). A control case is picked with $h=120$ nm, $p=390$ nm, and $d=0.6\,p$. Fig.~S14a shows the effect of varying $h$ from 100 nm to 400 nm in the control case. This figure shows that few values of $h$ give sharp reflections.  Increasing the height past 100 nm causes a redshift in the reflection and a severe broadening of the reflection spectrum, until it decreases around $h=300$ nm and ultimately disappears around $h=400$ nm. Also, around $h=200$ nm, another reflection appears in the spectrum. Increasing $h$ beyond this point causes a redshift without the same severe broadening.

Fig.~S14b shows the effect of varying $p$ from 200 nm to 500 nm in the control case. Fig.~S14b shows that increasing $p$ causes a redshift in the reflection spectrum throughout this test case. Also, the reflected spectrum narrows by increasing $p$ from around $p=200$ nm to around $p=400$ nm. Fig.~14c shows the effect of varying $d$ from $d=150$ nm to $d=350$ nm. Figure S8c shows that increasing $d$ from 150 nm causes a redshift in the resulting spectrum. This peak decreases for $d>250$ nm. However, another peak starts to appear around $d>250$ nm and remains at larger values if $d$ in this range. This new peak does not experience a red shift with an increase in $d$, but another, narrower, peak starts to appear with the increase in $d$. The change from amorphous to crystalline Sb$_2$S$_3$ has a nearly uniform effect in all these cases.  The phase change to crystalline severely decreases the reflectivity of the metasurface and causes a redshift at the same time. 

\begin{figure*}[htbp]
\centering
\includegraphics[page=1,width=0.85\linewidth, trim={0cm 0cm 0cm 0cm},clip]{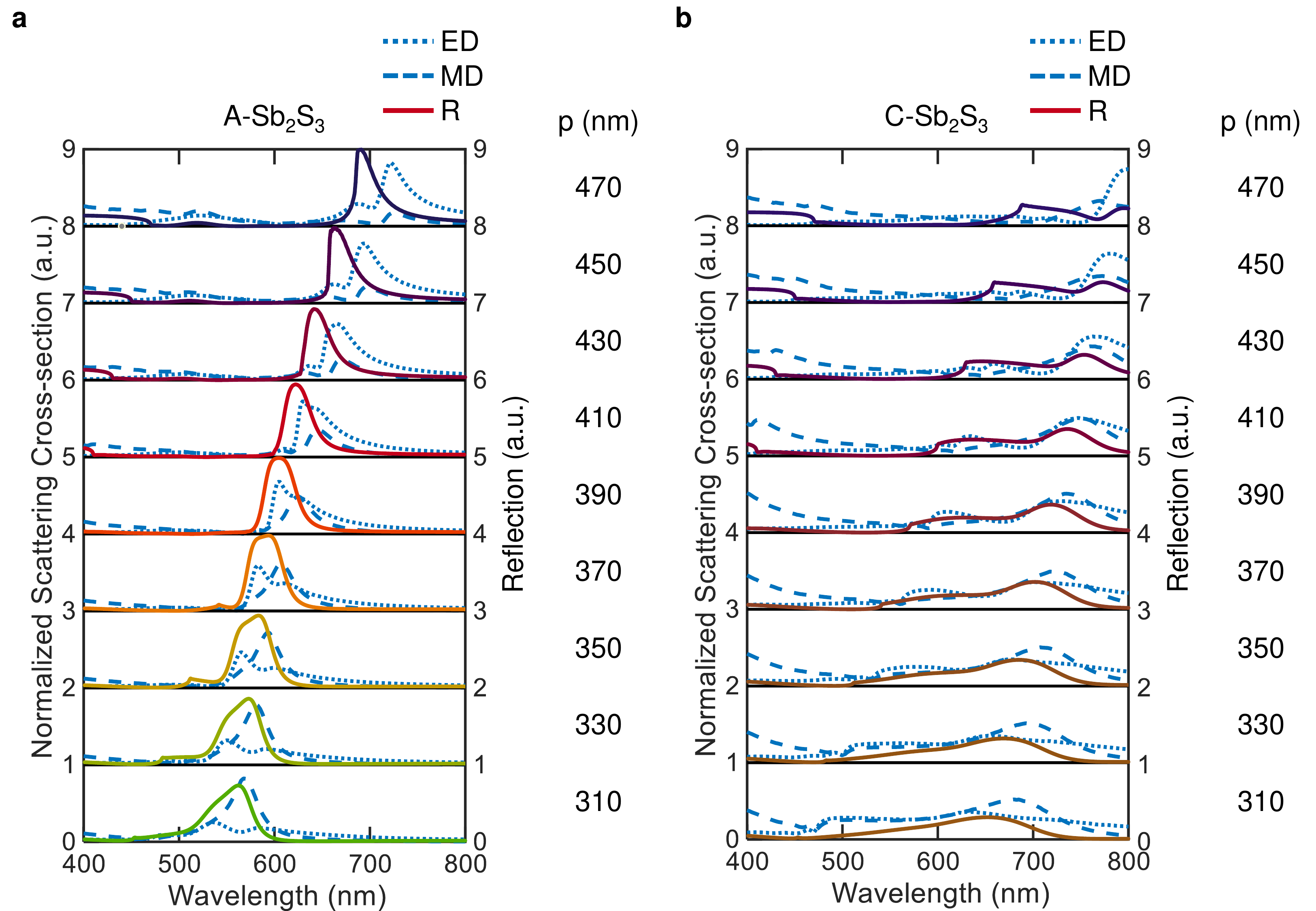}
\caption{\textbf{Multipolar decomposition  analysis.} \textbf{a},\textbf{b}, Multipolar decomposition of scattering cross-section in terms of electric dipole (ED, the dotted lines) and magnetic dipole (MD, the dashed lines) for the case of an periodic array of (\textbf{a}) amorphous and (\textbf{b}) crystalline Sb$_2$S$_3$ nanopillars with $h=120$ nm, $d=0.6 \, p$ in a lattice with varying periodicity of $p$ on top of a SiO$_2$ substrate. The reflectance (R) response for each case is plotted in solid lines.}
\label{Fig_S2}
\end{figure*}

\begin{figure*}[htbp]
\centering
\includegraphics[page=1,width=0.8\linewidth, trim={0cm 0cm 0cm 0cm},clip]{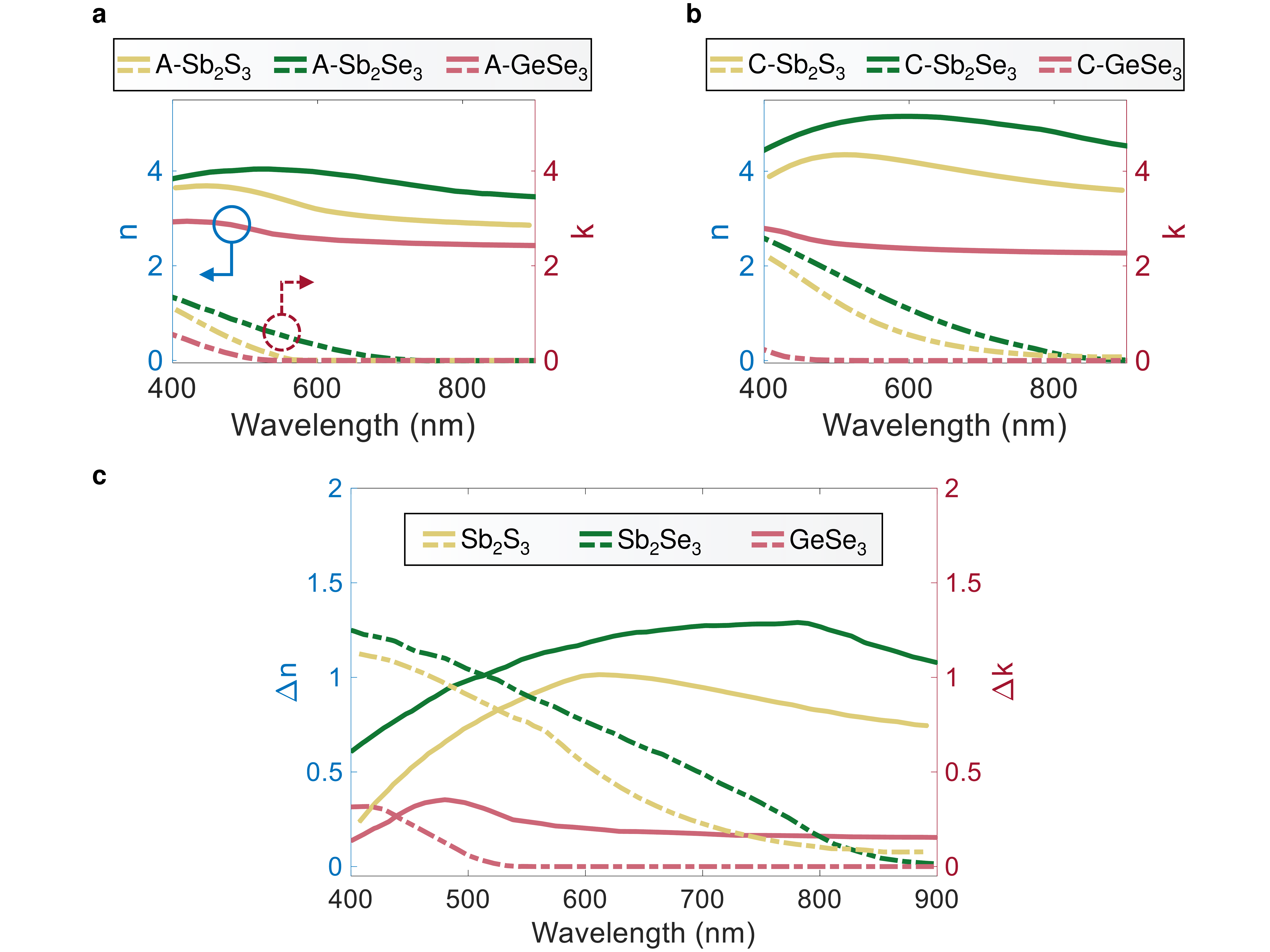}
\caption{\textbf{Optical characteristics of low-loss phase-change materials}. \textbf{a},\textbf{b}, Real (solid lines) and imaginary (dashed lines) parts of the refractive index of Sb$_2$S$_3$, Sb$_2$Se$_3$, and GeSe$_3$ for (\textbf{a}) amorphous (A) and (\textbf{b}) crystalline (C) phases. \textbf{c}, The absolute value of the change in the refractive index (solid lines, $\Delta n = |n_{\textrm{C-PCM}} - n_{\textrm{A-PCM}}|$) and the extinction coefficient (dashed lines, $\Delta k = |k_{\textrm{C-PCM}} - k_{\textrm{A-PCM}}|$) versus the wavelength upon the transition between amorphous and crystalline phase-states for Sb$_2$S$_3$, Sb$_2$Se$_3$ and GeSe$_3$.}
\label{Fig_S1}
\end{figure*}

\begin{figure*}[htbp]
\centering
\includegraphics[width=.65\linewidth, trim={0cm 0cm 0cm 0cm},clip]{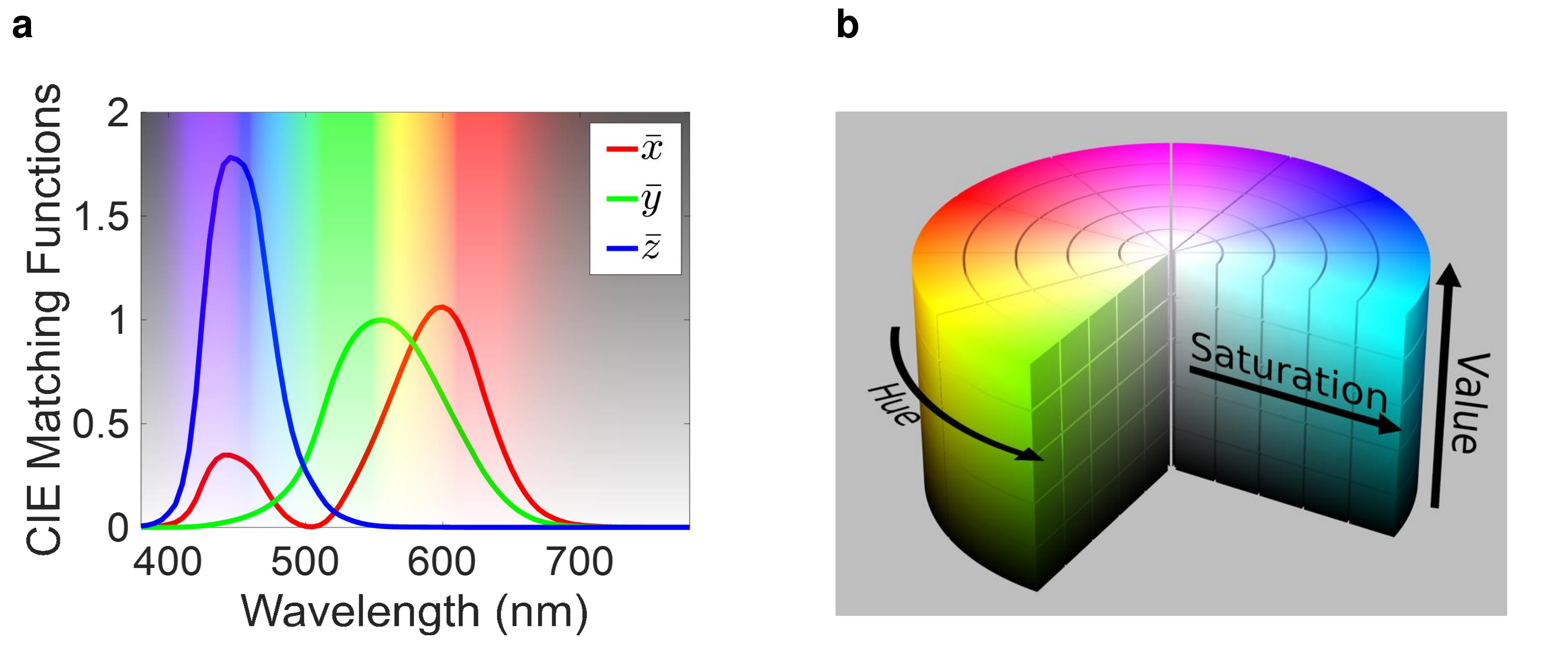}
\caption{\textbf{Color generation and characteristics.} \textbf{a}, CIE 1931
standard color-matching functions. \textbf{b}, HSV color solid cylinder saturation gray \cite{hemmatyar2019full}.}
\label{Fig_S2}
\end{figure*}

\begin{figure*}[htbp]
\centering
\includegraphics[width=1\linewidth, trim={0cm 0cm 0cm 0cm},clip]{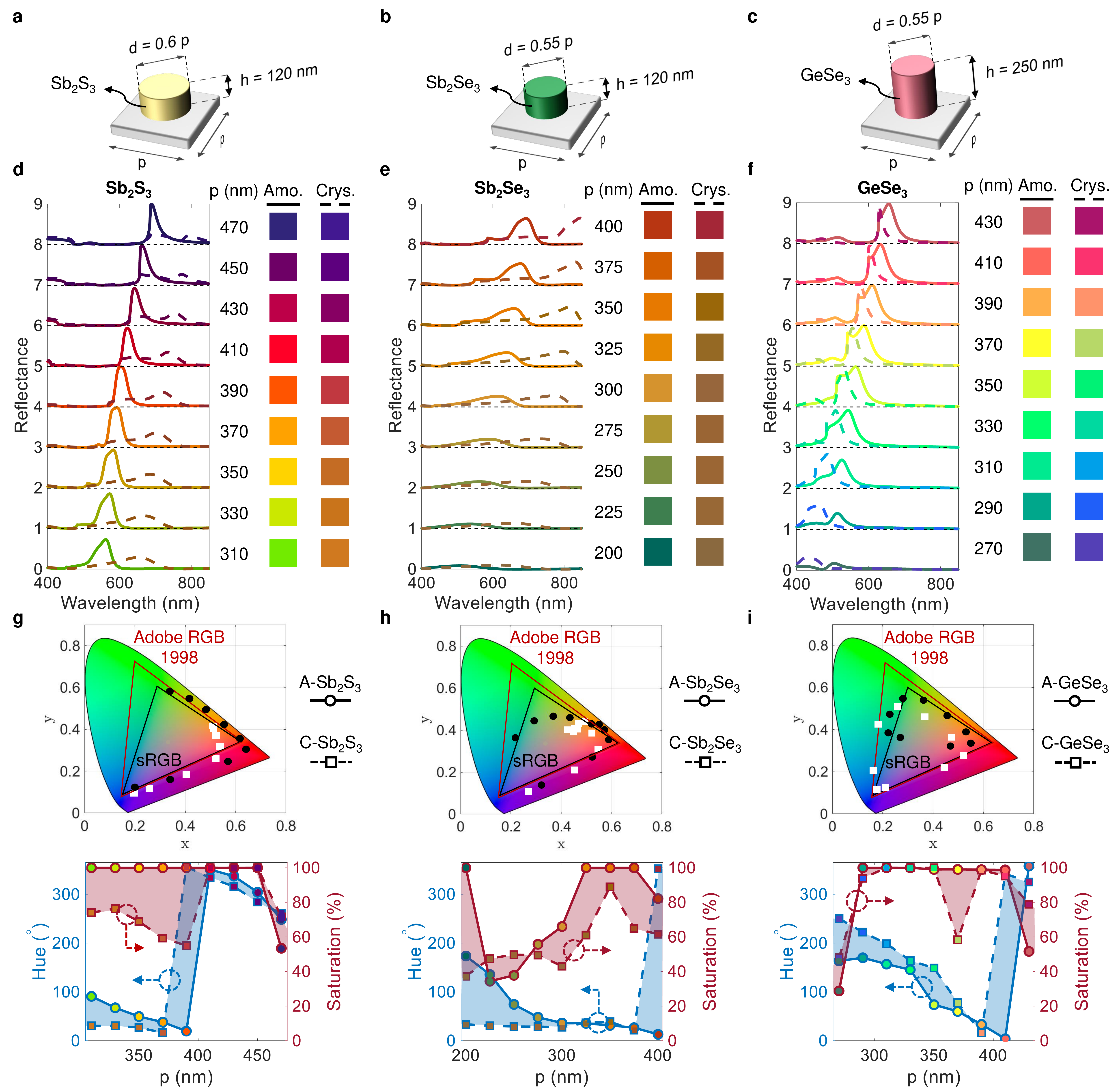}
\caption{\textbf{Color switching enabled by phase-transition of the PCM nanopillars.} \textbf{a-c}, Schematic and geometrical parameters of a unit cell of a polarization-insensitive PCM metasurface made of (\textbf{a}) Sb$_2$S$_3$, (\textbf{b}) Sb$_2$Se$_3$ and (\textbf{c}) GeSe$_3$ circular nanopillars with a fixed heigh $h$. The periodicity of the unit cell in both x and y directions is $p$, and the diameter of the nanopillars is $d = \alpha \, p$ with $\alpha$ being a constant. \textbf{d-f}, Simulated reflectance spectra for the amorphous (solid lines) and crystalline (dashed lines) phases and their corresponding colors for different periodicities ($p$). The PCM is (\textbf{d}), (\textbf{e}), and (\textbf{f}) is Sb$_2$S$_3$, Sb$_2$Se$_3$, and GeSe$_3$, respectively. The curves for different $p$s are diplaced vertically for better visibility and comparison. The sharp resonances observed in (\textbf{d-f}) are attributed to the interference between ED and MD modes inside the PCM nanopillars. Upon the PCM phase transition, a red-shift of $|\Delta\lambda_{\textrm{Sb}_{2}\textrm{S}_{3}}|>180$ nm and $|\Delta\lambda_{\textrm{Sb}_{2}\textrm{Se}_{3}}|>200$ nm is observed for the case of (\textbf{d}) Sb$_2$S$_3$ and (\textbf{e}) Sb$_2$Se$_3$, respectively, while a blue-shift of $|\Delta\lambda_{\textrm{Ge}\textrm{Se}_{3}}|<70$ nm is observed for the case of (\textbf{f}) GeSe$_3$. \textbf{g-i}, Corresponding CIE 1931 chromaticity coordinates of the reflectance spectra, and the hue and saturation values of the colors shown in (\textbf{d-f}) for amorphous (black circles in the top panel and circle-solid line in bottom panel) and crystalline (white squares in the top panel and square-dashed line in the bottom panel) phases of the corresponding PCMs in (\textbf{d-f}).}
\label{Fig_2}
\end{figure*}

\begin{figure*}[htbp]
\centering
\includegraphics[width=.55\linewidth, trim={0cm 0cm 0cm 0cm},clip]{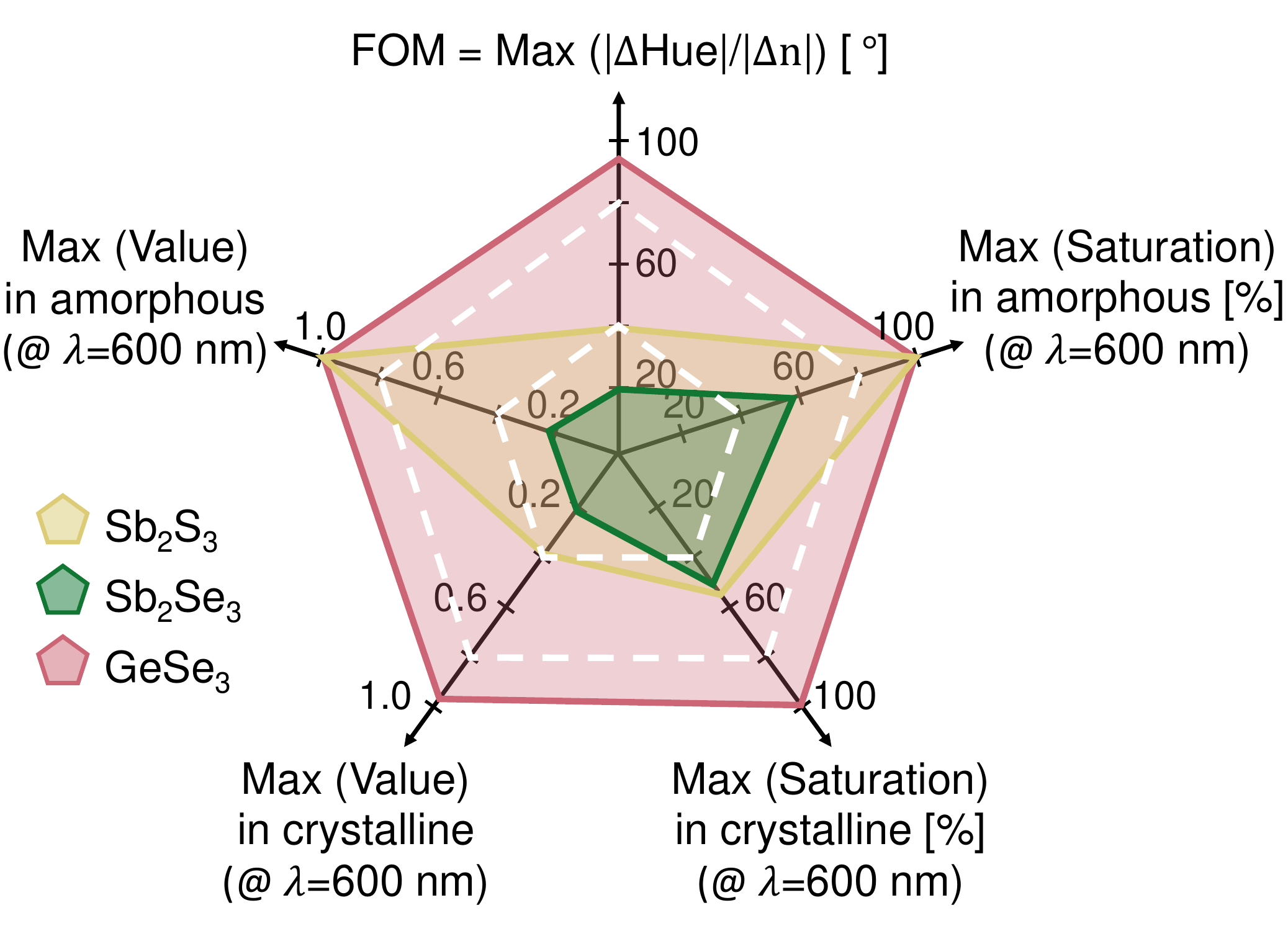}
\caption{\textbf{Comparison of low-loss PCMs for color switching applications.} A spider chart that compares Sb$_2$S$_3$, Sb$_2$Se$_3$ and GeSe$_3$ in terms of FOM (defined as the maximum (max) of $|\Delta \textrm{Hue}|/|\Delta n|$ in which $\Delta n = n_\textrm{A}(\lambda_\textrm{A}) - n_\textrm{C}(\lambda_\textrm{C})$), maximum saturation and maximum value (i.e. the reflectance value at the resonance peak) in amorphous and crystalline phases at $\lambda_\textrm{A} =600$~nm.}
\label{Fig_6}
\end{figure*}

\begin{figure*}[htbp]
\centering
\includegraphics[width=1\linewidth, trim={0cm 0cm 0cm 0cm},clip]{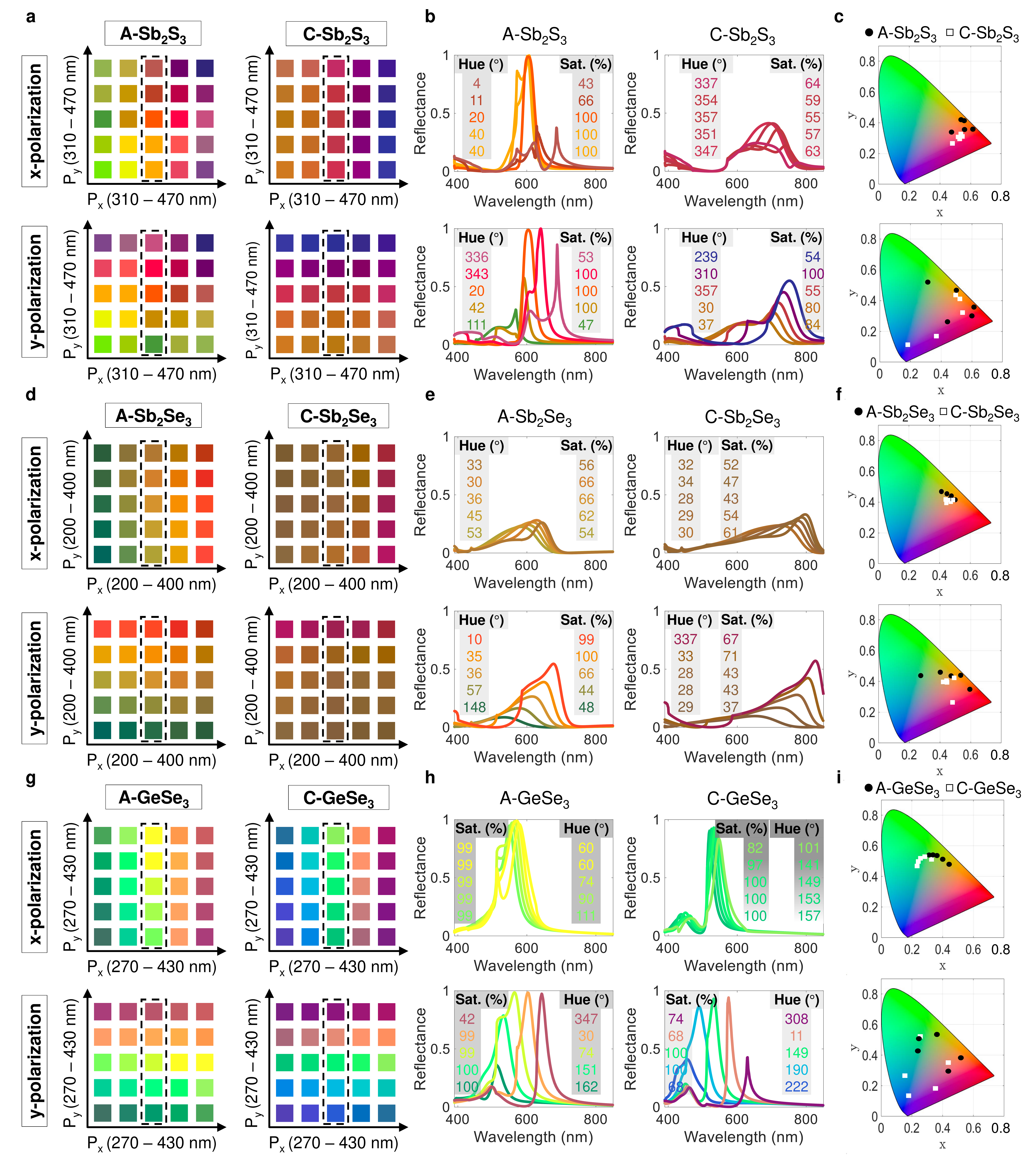}
\caption{\textbf{Multiple color generation enabled by phase-transition-based and polarization-based color switching mechanisms.} \textbf{a,d,g}, Generated color palettes considering different periodicities in x- and y-directions ($p_x$ and $p_y$, respectively) for (\textbf{a}) Sb$_2$S$_3$ ($\alpha=0.6$ and $h=120$ nm), (\textbf{d}) Sb$_2$Se$_3$ ($\alpha=0.55$ and $h=120$ nm), and (\textbf{g}) GeSe$_3$ ($\alpha=0.55$ and $h=250$ nm). $p_x$ and $p_y$ in (\textbf{a}), (\textbf{d}) and (\textbf{g}) vary with 40 nm, 50 nm, and 40 nm increments, respectively. \textbf{b,e,h}, Reflectance spectra of the colors indicated by the dashed rectangular boxes shown in the corresponding color palette in (\textbf{a,d,g}), respectively, with the values of hue and saturation (sat.) in the inset. \textbf{c,f,i}, Corresponding color gamuts for amorphous (black circles) and crystalline (white squares) phases of the corresponding PCM in (\textbf{a,d,g}), respectively. In each figure, the upper (lower) panel represents the results related to x-polarization (y-polarization).}
\label{Fig_6}
\end{figure*}

\begin{figure*}[htbp]
\centering
\includegraphics[page=1,width=0.8\linewidth, trim={0cm 0cm 0cm 0cm},clip]{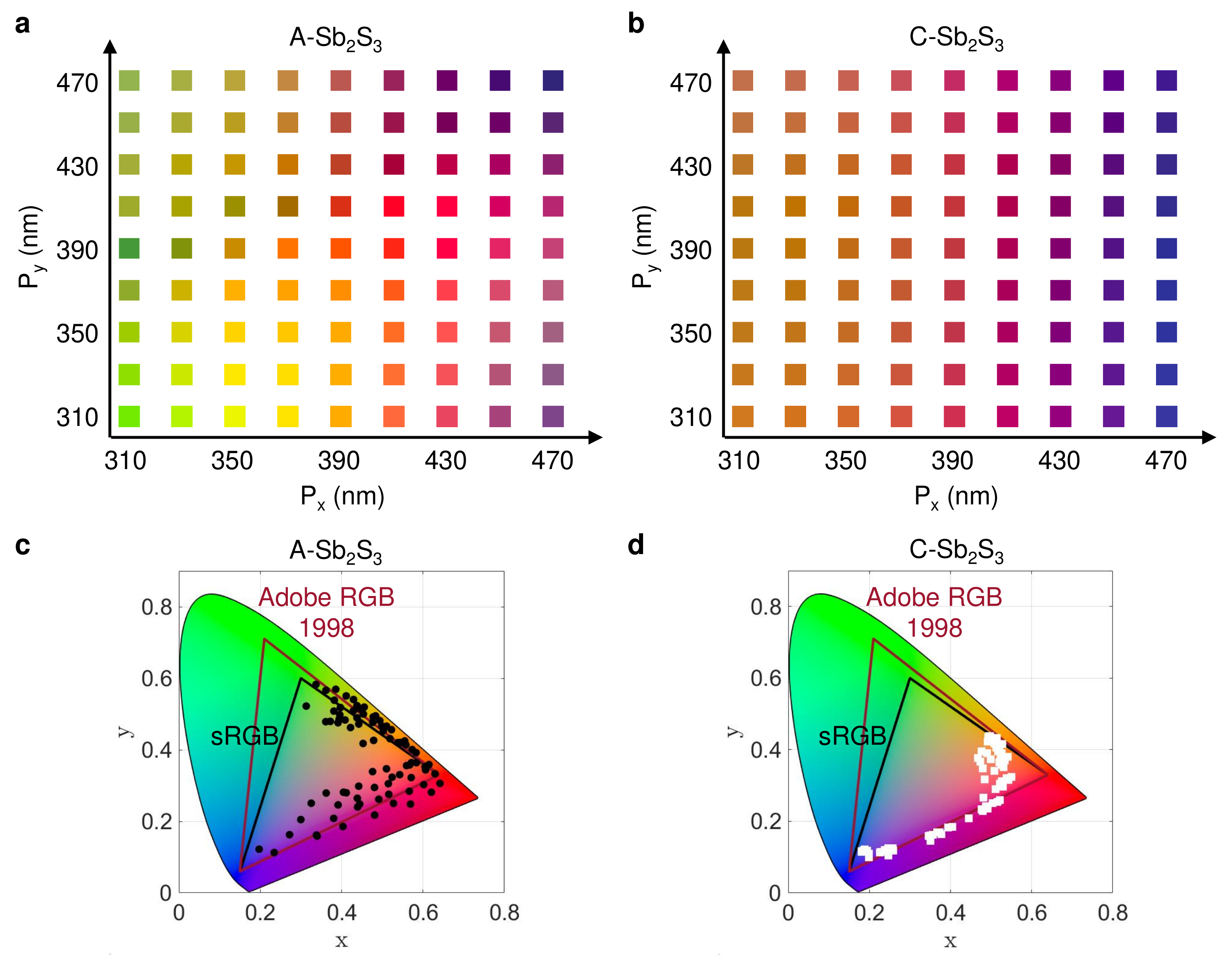}
\caption{\textbf{Dynamic color generation by Sb$_2$S$_3$ meta-pixels.} \textbf{a,b}, The color palettes and \textbf{c, d}, corresponding CIE 1931 chromaticity diagrams generated by Sb$_2$S$_3$ metasurfaces in (\textbf{a, c}) amorphous and (\textbf{b, d}) crystalline phase-states under x-polarized normally incident white light. The lattice periodicities in x- and y-directions vary from $p_{\textrm{x,y}} = 310$ nm to $p_{\textrm{x,y}}  = 470$ nm  with a step of 20 nm while the diameter of the nanopillars changes as $d_{\textrm{x,y}} =0.6 \, p_{\textrm{x,y}} $, and the height of the nanopillars is fixed at $h=120$ nm.}
\label{Fig_S7}
\end{figure*}

\begin{figure*}[htbp]
\centering
\includegraphics[page=1,width=0.8\linewidth, trim={0cm 0cm 0cm 0cm},clip]{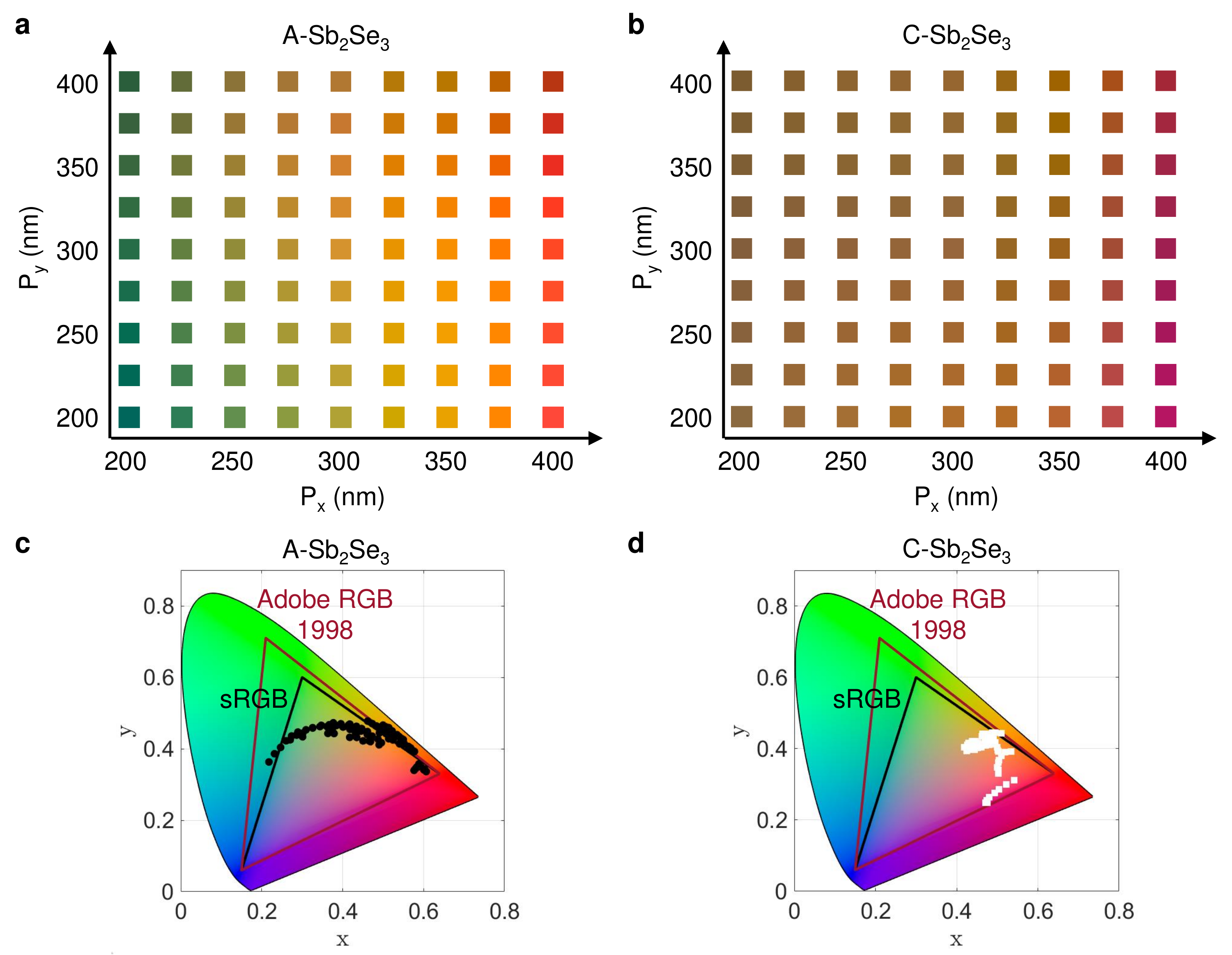}
\caption{\textbf{Dynamic color generation by Sb$_2$Se$_3$ meta-pixels.} \textbf{a,b}, The color palettes and \textbf{c, d}, corresponding CIE 1931 chromaticity diagrams generated by Sb$_2$Se$_3$ metasurfaces in (\textbf{a, c}) amorphous and (\textbf{b, d}) crystalline phase-states under x-polarized normally incident white light. The lattice periodicities in x- and y-directions vary from $p_{\textrm{x,y}} = 200$ nm to $p_{\textrm{x,y}}  = 400$ nm  with a step of 25 nm while the diameter of the nanopillars changes as $d_{\textrm{x,y}} =0.55 \, p_{\textrm{x,y}} $, and the height of the nanopillars is fixed at $h=120$ nm.}
\label{Fig_S8}
\end{figure*}

\begin{figure*}[htbp]
\centering
\includegraphics[page=1,width=0.8\linewidth, trim={0cm 0cm 0cm 0cm},clip]{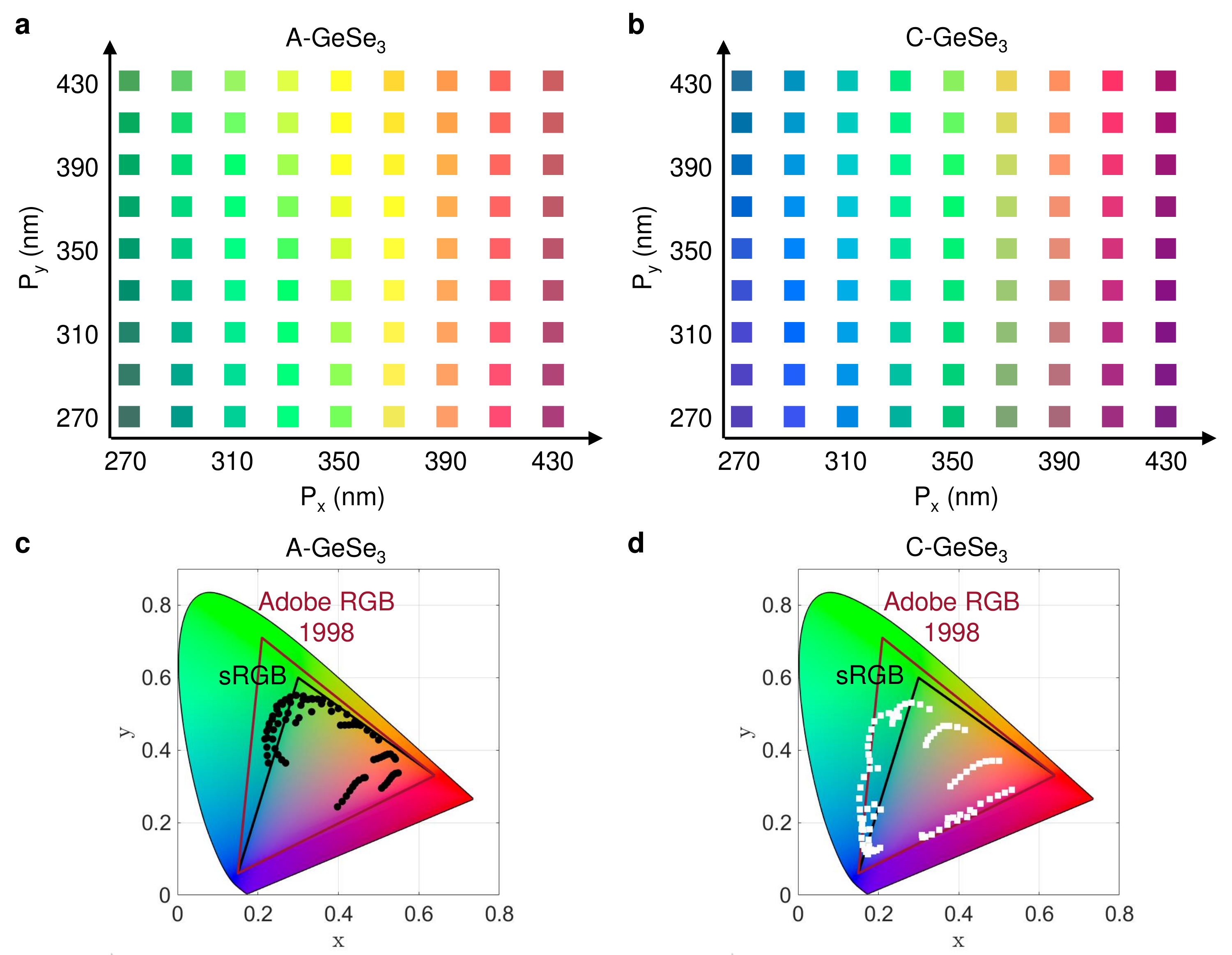}
\caption{\textbf{Dynamic color generation by GeSe$_3$ meta-pixels.} \textbf{a,b}, The color palettes and \textbf{c, d}, corresponding CIE 1931 chromaticity diagrams generated by GeSe$_3$ metasurfaces in (\textbf{a, c}) amorphous and (\textbf{b, d}) crystalline phase-states under x-polarized normally incident white light. The lattice periodicities in x- and y-directions vary from $p_{\textrm{x,y}} = 270$ nm to $p_{\textrm{x,y}}  = 430$ nm  with a step of 20 nm while the diameter of the nanopillars changes as $d_{\textrm{x,y}} =0.55 \, p_{\textrm{x,y}} $, and the height of the nanopillars is fixed at $h=250$ nm.}
\label{Fig_S9}
\end{figure*}

\begin{figure*}[htbp]
\centering
\includegraphics[page=1,width=0.9\linewidth, trim={0cm 0cm 0cm 0cm},clip]{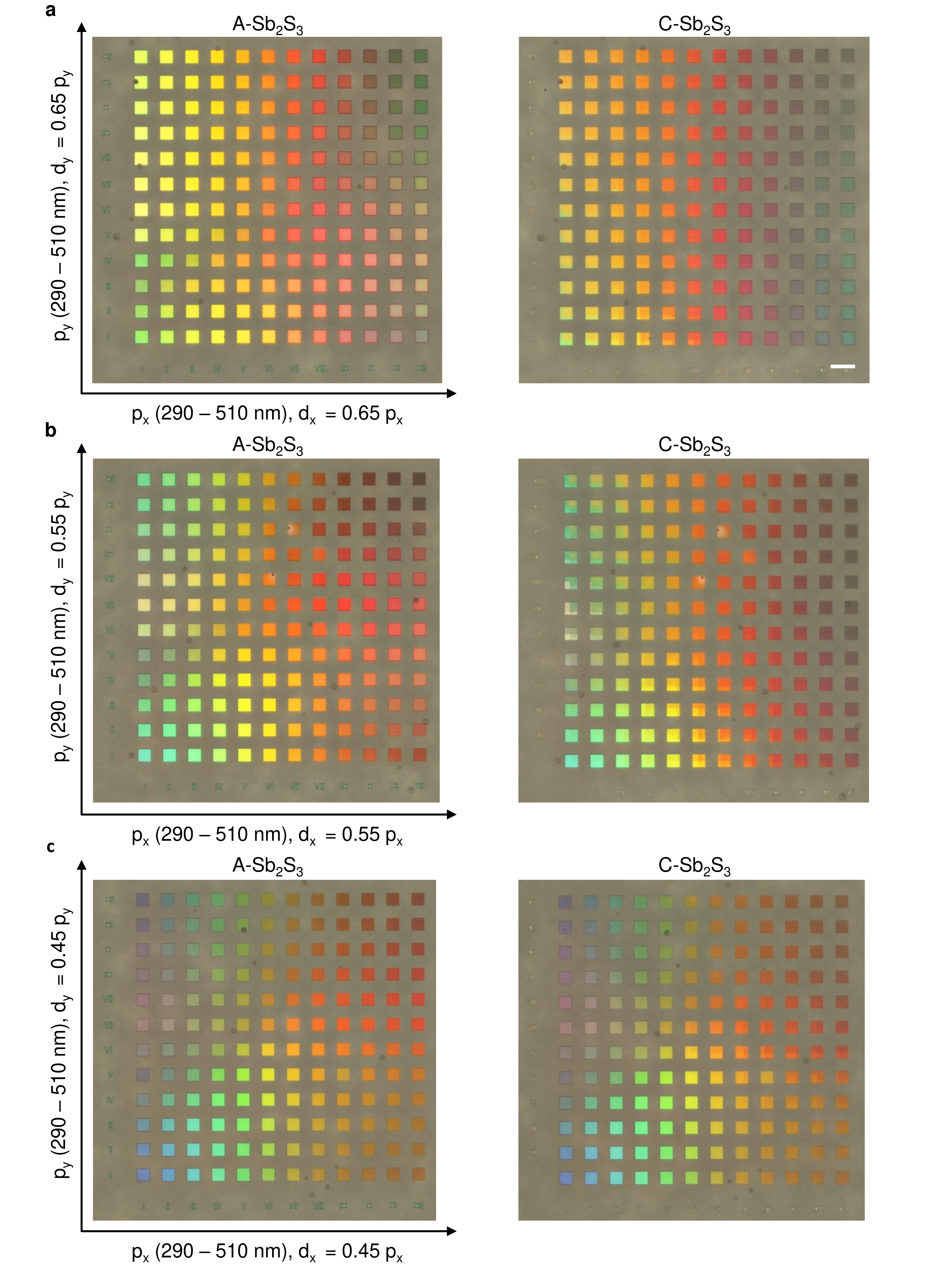}
\caption{\textbf{Experimental color palettes of Sb$_2$S$_3$ meta-pixels} \textbf{a,b,c}, A-Sb$_2$S$_3$ (left) and C-Sb$_2$S$_3$ (right) meta-pixels considering different periodicities in x- and y- directions ($p_x$ and $p_y$, respectively) varying with 20 nm increments while the diameter of the nanopillars changes as (\textbf{a}) $d_{\textrm{x,y}} =0.65 \, p_{\textrm{x,y}}$, (\textbf{b}) $d_{\textrm{x,y}} =0.55 \, p_{\textrm{x,y}}$, (\textbf{c}) $d_{\textrm{x,y}} =0.45 \, p_{\textrm{x,y}}$, and the height of the nanopillars is fixed at $h=120$ nm. }
\label{Fig_S10}
\end{figure*}

\begin{figure*}[htbp]
\centering
\includegraphics[page=1,width=0.9\linewidth, trim={0cm 0cm 0cm 0cm},clip]{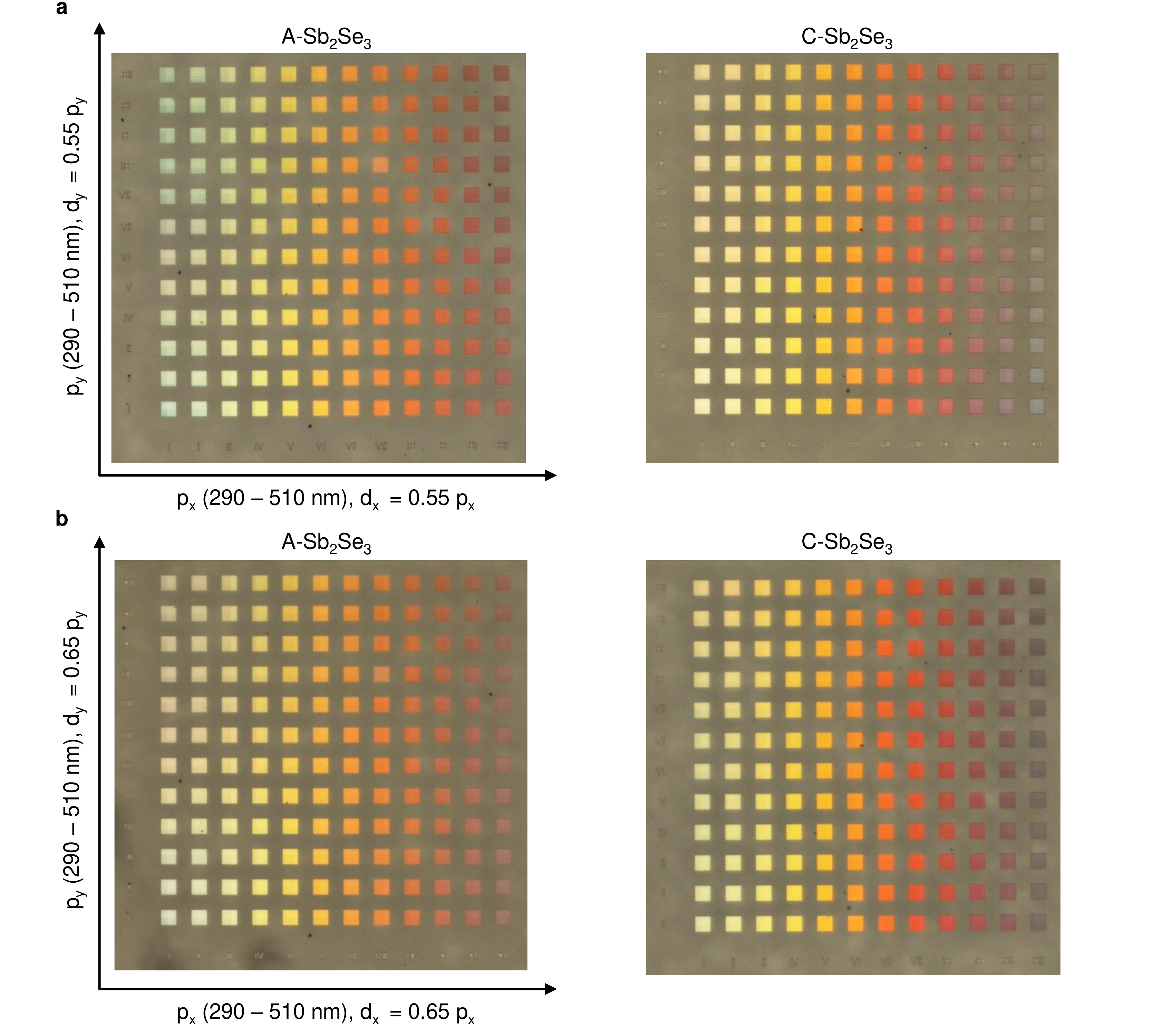}
\caption{\textbf{Experimental color palettes of Sb$_2$Se$_3$ meta-pixels} \textbf{a,b}, A-Sb$_2$Se$_3$ (left) and C-Sb$_2$Se$_3$ (right) meta-pixels considering different periodicities in x- and y- directions ($p_x$ and $p_y$, respectively) varying with 20 nm increments while the diameter of the nanopillars changes as (\textbf{a}) $d_{\textrm{x,y}} =0.55 \, p_{\textrm{x,y}}$, (\textbf{b}) $d_{\textrm{x,y}} =0.65 \, p_{\textrm{x,y}}$, and the height of the nanopillars is fixed at $h=120$ nm. Sb$_2$Se$_3$ is sputtered in a magnetron sputtering system using 30 W radio frequency (RF) power at a deposition pressure of 4 mTorr and Ar flow of 30 sccm. The deposition rate for Sb$_2$Se$_3$ is $\sim$1 nm/min. Before deposition, the chamber base pressure is maintained at $\sim$10$^{-7}$ Torr. Additionally, the samples are capped with 15 nm of SiO$_2$ sputtered in situ, to prevent oxidation during later characterization. As an aside, several pre- and post-deposition treatments of the sputtering chamber are performed for selenide deposition. These include cleaning the chamber followed by annealing and O$_2$ plasma cleaning.}
\label{Fig_S11}
\end{figure*}

\begin{figure*}[htbp]
\centering
\includegraphics[width=.95\linewidth, trim={0cm 0cm 0cm 0cm},clip]{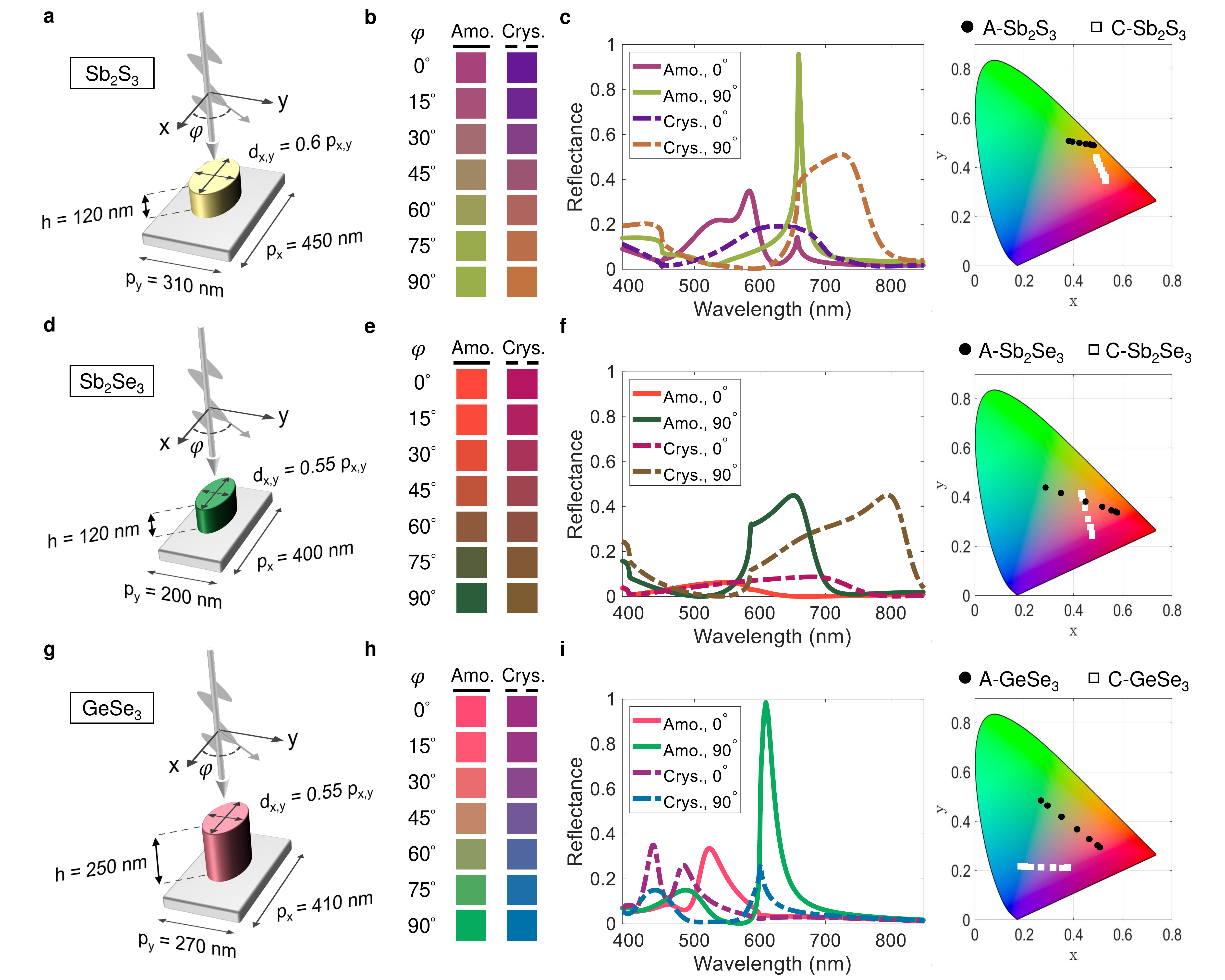}
\caption{\textbf{Polarization-based continuous color-switching enabled by rotating the incident polarization angle.} The asymmetric unit cells of the polarization-sensitive metasurface with the optimized design parameters are shown in \textbf{a}, \textbf{d}, and \textbf{g}, respectively, with their corresponding variation of colors with polarization angle $\varphi$ and their color gamts shown in \textbf{b}, \textbf{e}, and \textbf{h}, respectively. The simulated reflectance spectra from \textbf{c}, Sb$_2$S$_3$, \textbf{f}, Sb$_2$Se$_3$, and \textbf{i}, GeSe$_3$ metasurfaces for x-polarization ($\varphi = 90^\circ$) and y-polarization ($\varphi = 0^\circ$). The reflection-mode color response varies from reddish purple to the yellowish green for A-Sb$_2$S$_3$, bluish purple to the reddish orange for C-Sb$_2$S$_3$, red to dark green for A-Sb$_2$Se$_3$, red purple to brown for C-Sb$_2$Se$_3$, purple to green for A-GeSe$_3$, and red purple to blue for C-GeSe$_3$.}
\label{Fig_S12}
\end{figure*}

\begin{figure*}[htbp]
\centering
\includegraphics[page=1,width=0.55\linewidth, trim={0cm 0cm 0cm 0cm},clip]{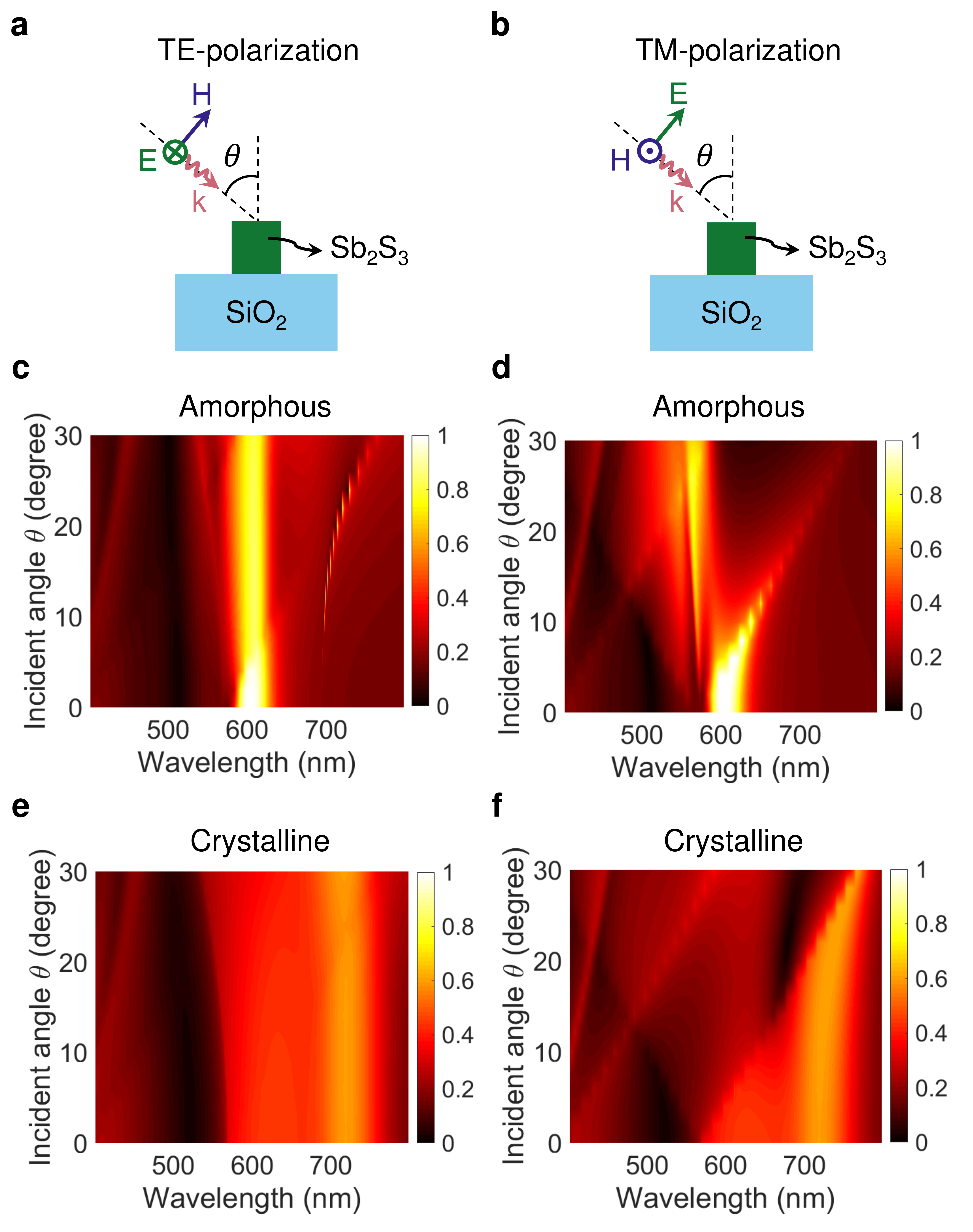}
\caption{\textbf{Analysis of the sensitivity to the angle of incidence.} The structure used in the study of the angle sensitivity of a Sb$_2$S$_3$ metasurface for the case of obliquely incident plane waves of white light for \textbf{a}, TE and \textbf{b}, TM polarizations, respectively. \textbf{c,d,e,f}, The simulated reflection spectra of the metasurface, showing the incident angle (degrees) versus wavelength (nm) for: (\textbf{c}) amorphous phase and TE polarization, (\textbf{d}) amorphous phase and TM polarization, (\textbf{e}) crystalline phase and TE polarization, (\textbf{f}) crystalline phase and TM polarization.}
\label{Fig_S13}
\end{figure*}

\begin{figure*}[b]
\centering
\includegraphics[page=1,width=0.85\linewidth, trim={0cm 0cm 0cm 0cm},clip]{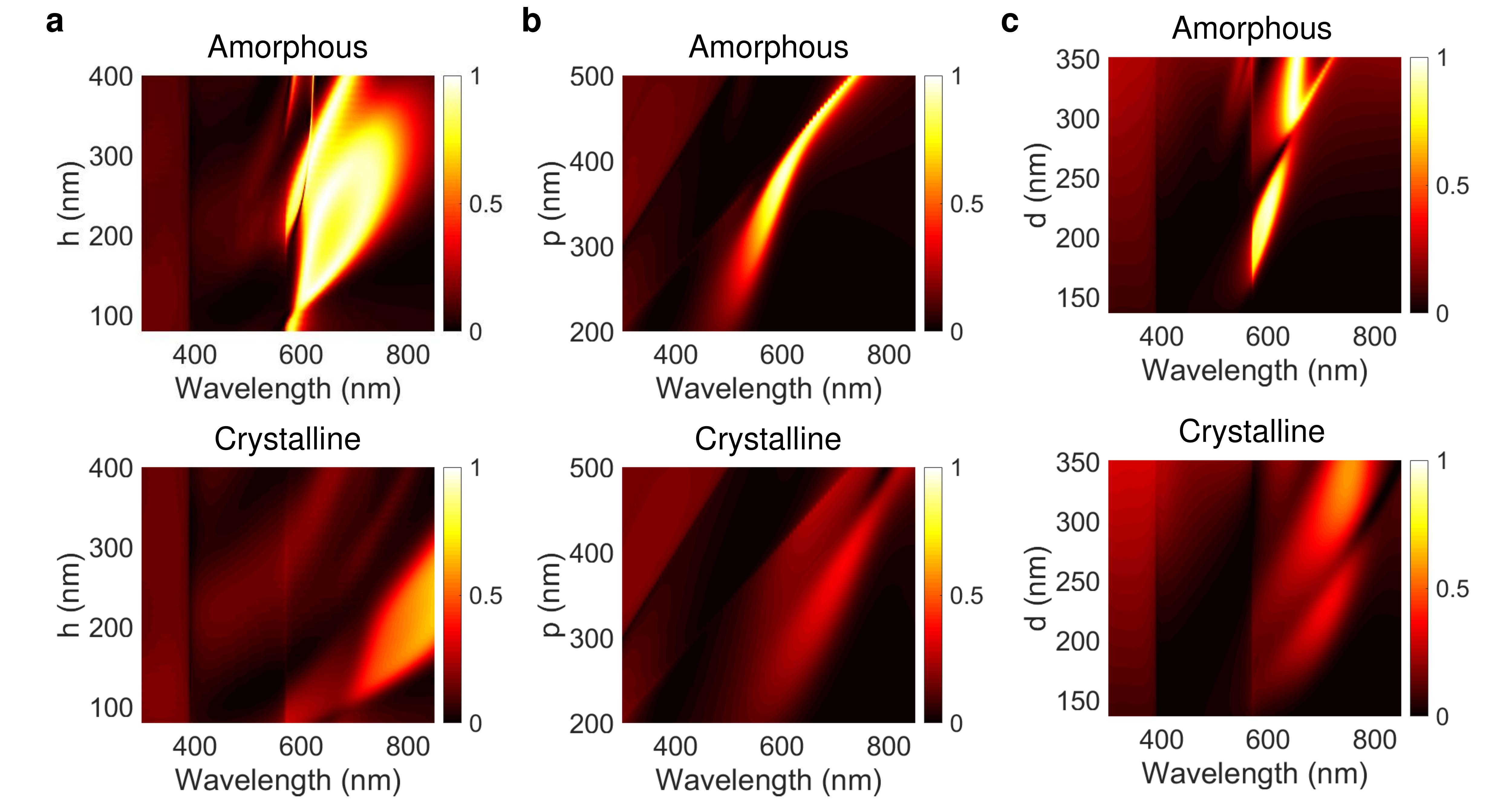}
\caption{\textbf{Analysis of the effect of different design parameters.} Simulated reflection spectrum of the Sb$_2$S$_3$ metasurface in Figure~2a versus; \textbf{a}, The height of the constituents nanopillars, i.e., $h$, while other parameters are fixed at $p = 390$ nm and $d = 0.6\,p$ for (top) amorphous and (bottom) crystalline phases; \textbf{b}, period of the unit cell, i.e., $p$, with $d = 0.6\,p$ and $h = 120$ nm for (top) amorphous and (bottom) crystalline phases; \textbf{c}, diameter of the constituent nanopillars, i.e., $d$, with $p = 390$ nm and $h = 120$ nm  for (top) amorphous and (bottom) crystalline phases.}
\label{Fig_S14}
\end{figure*}

\begin{figure*}[htbp]
\centering
\includegraphics[page=1,width=1\linewidth, trim={0cm 0cm 0cm 0cm},clip]{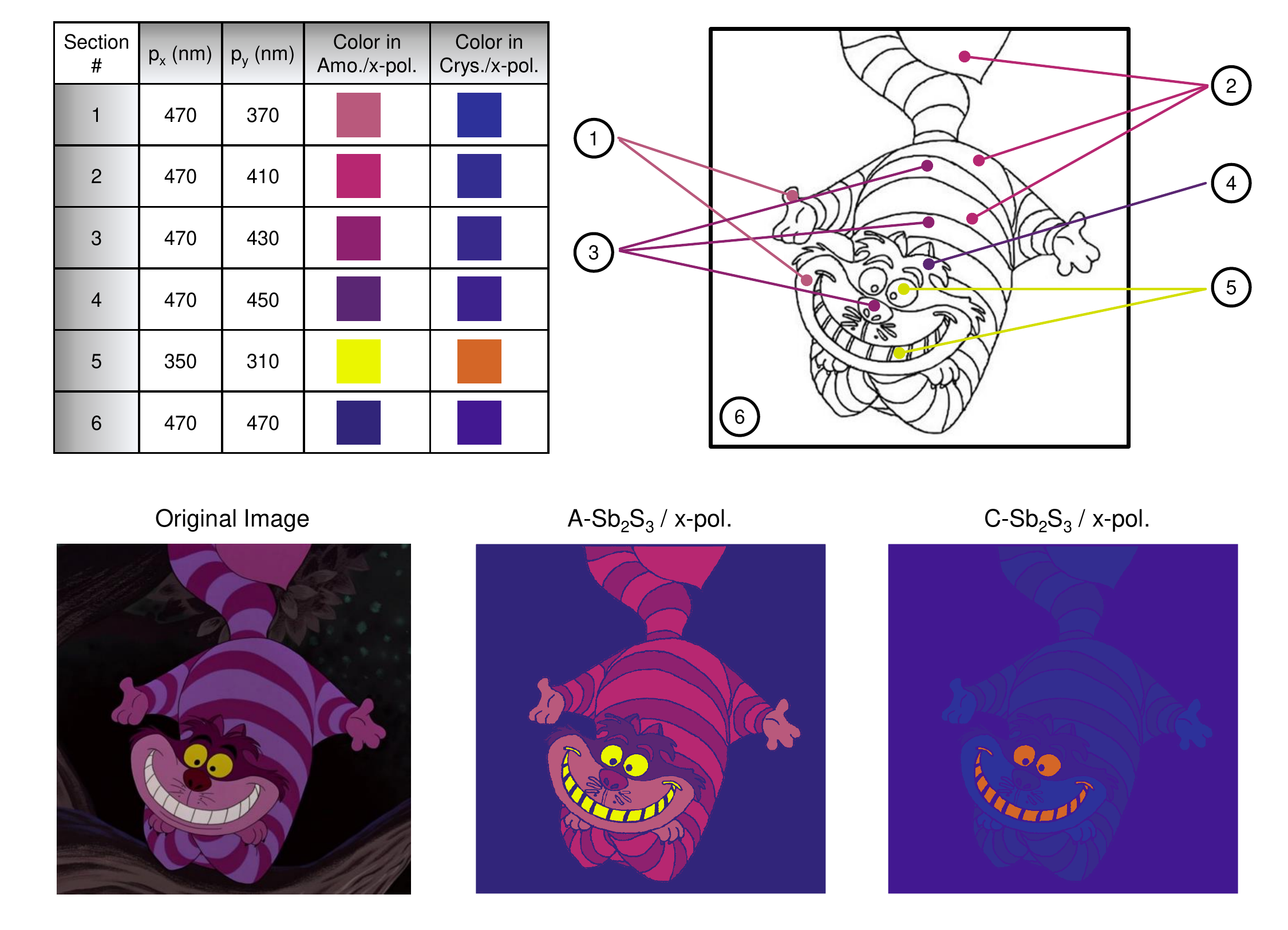}
\caption{\textbf{Design strategy for generating the dynamic image of Cheshire The Cat.} The geometrical parameters of the Sb$_2$S$_3$ metasurfaces used for producing each pixel of the image of Cheshire The Cat shown in Fig.~3a in the main text ($d_{\textrm{x,y}} =0.65 \, p_{\textrm{x,y}}$ and $h=120$ nm in Fig.~1b).}
\label{Fig_S15}
\end{figure*}

\begin{figure*}[htbp]
\centering
\includegraphics[page=1,width=1\linewidth, trim={0cm 0cm 0cm 0cm},clip]{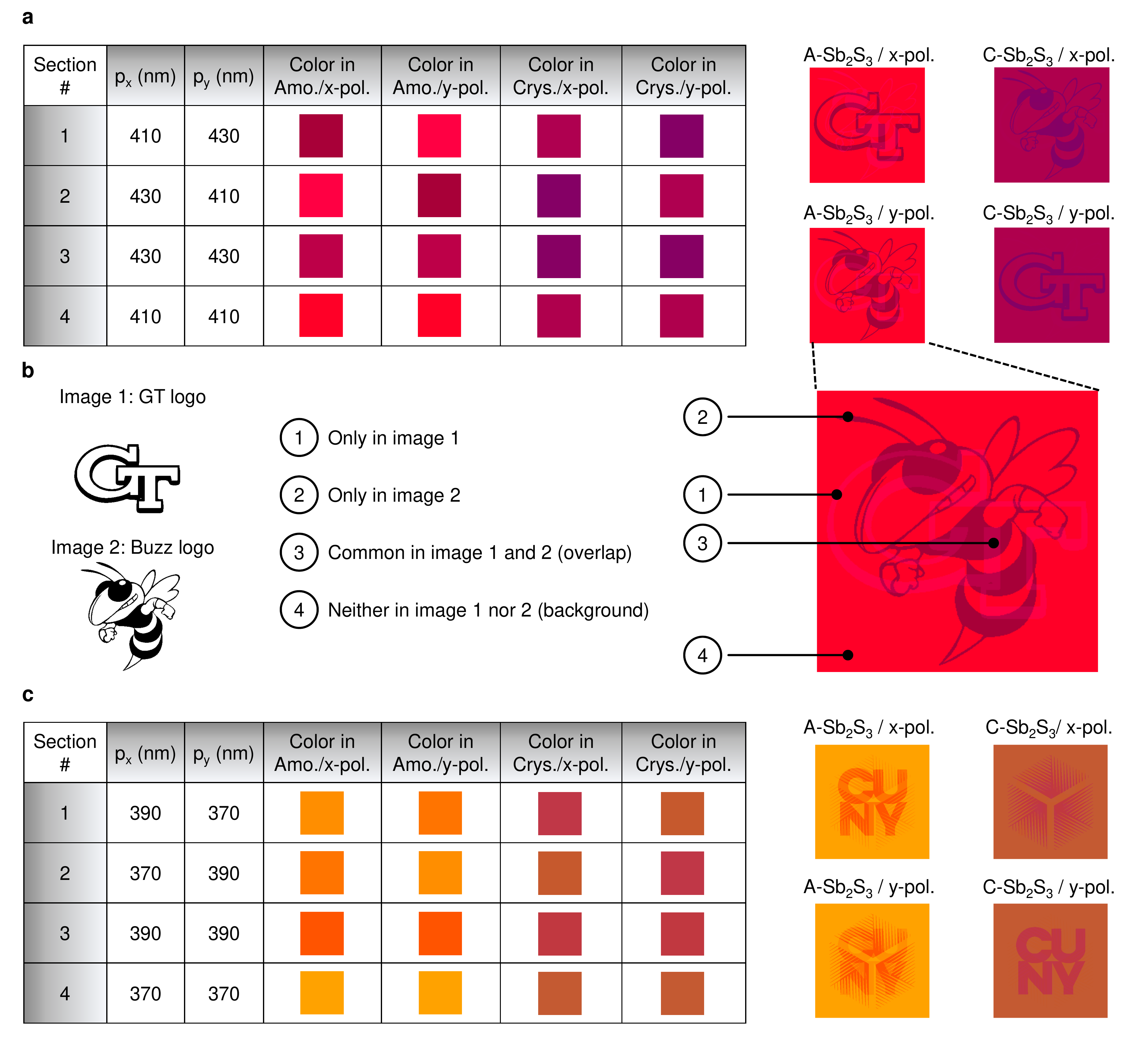}
\caption{\textbf{Design strategy for encryption of four different images into the phase and polarization of Sb$_2$S$_3$ meta-pixels.} \textbf{a,c}, Phase-transition-based switching between two different images. The colors are generated by four different metasurfaces consisting of Sb$_2$S$_3$ nanopillars (Fig.~1b) with periodicities reported in the table, diameters $d_{\textrm{x,y}} =0.65 \, p_{\textrm{x,y}}$, and a fixed height $h=120$ nm. \textbf{b}, The definition of different zones in each image.}
\label{Fig_S12}
\end{figure*}

\begin{figure*}[htbp]
\centering
\includegraphics[page=1,width=1\linewidth, trim={0cm 0cm 0cm 0cm},clip]{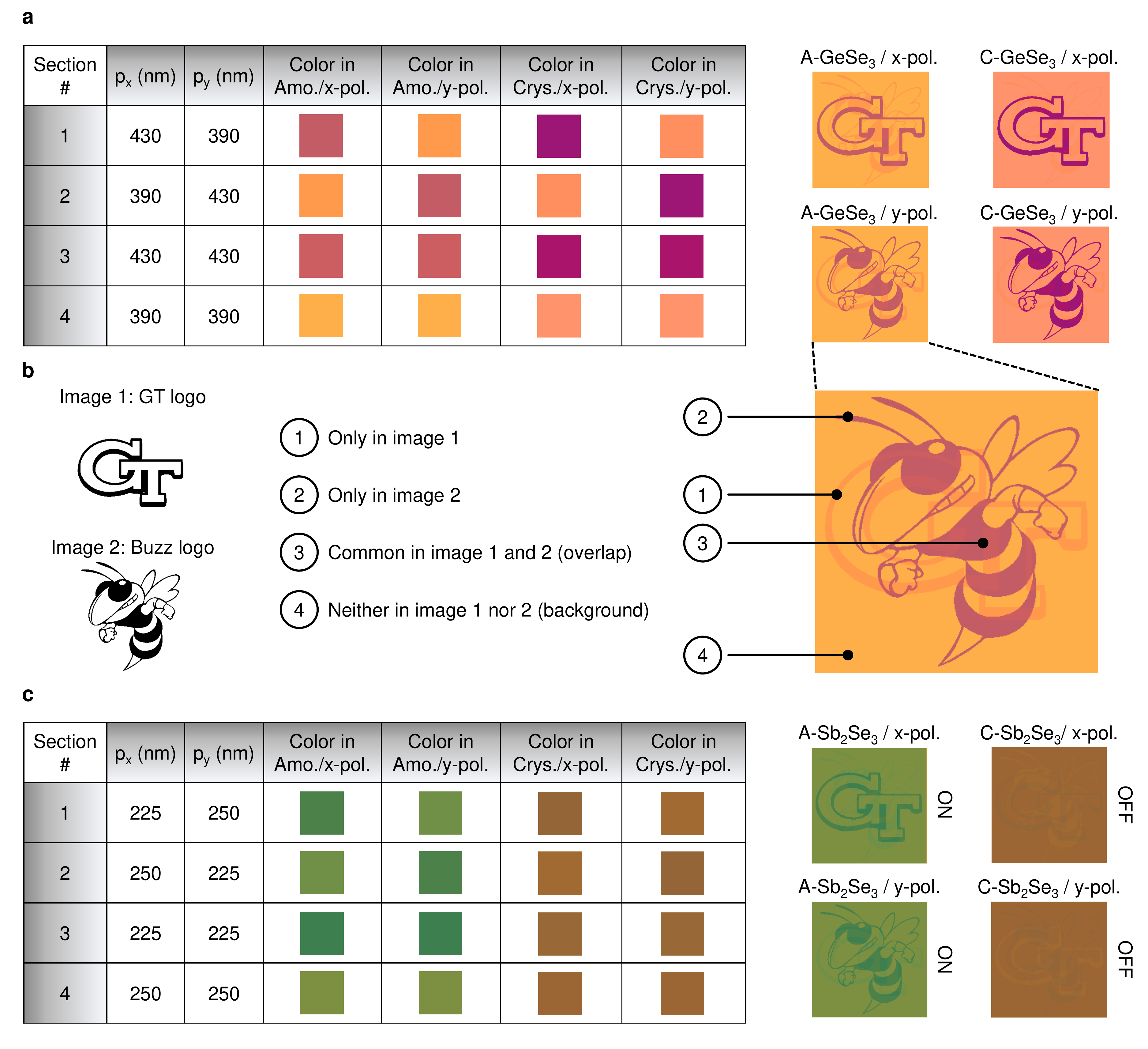}
\caption{\textbf{Design strategy for encryption of four different images into the phase and polarization of Sb$_2$Se$_3$ and GeSe$_3$ meta-pixels.} \textbf{a}, Polarization-based switching between two different images. The colors are generated by four different metasurfaces consisting of GeSe$_3$ nanopillars (Fig.~S3b) with periodicities reported in the table, diameters of $d_{\textrm{x,y}} =0.55 \, p_{\textrm{x,y}} $, and a fixed height $h=250$ nm. \textbf{b}, The definition of different zones in each image. \textbf{c}, Phase-transition-based switching between the ON-state (amorphous) and the OFF-state (crystalline). The colors are generated by four different metasurfaces consisting of Sb$_2$Se$_3$ nanopillars with periodicities reported in the table, diameters $d_{\textrm{x,y}} =0.55 \, p_{\textrm{x,y}} $, and a fixed height $h=120$ nm.}
\label{Fig_S12}
\end{figure*}

\begin{figure*}[htbp]
\centering
\includegraphics[page=1,width=1\linewidth, trim={0cm 0cm 0cm 0cm},clip]{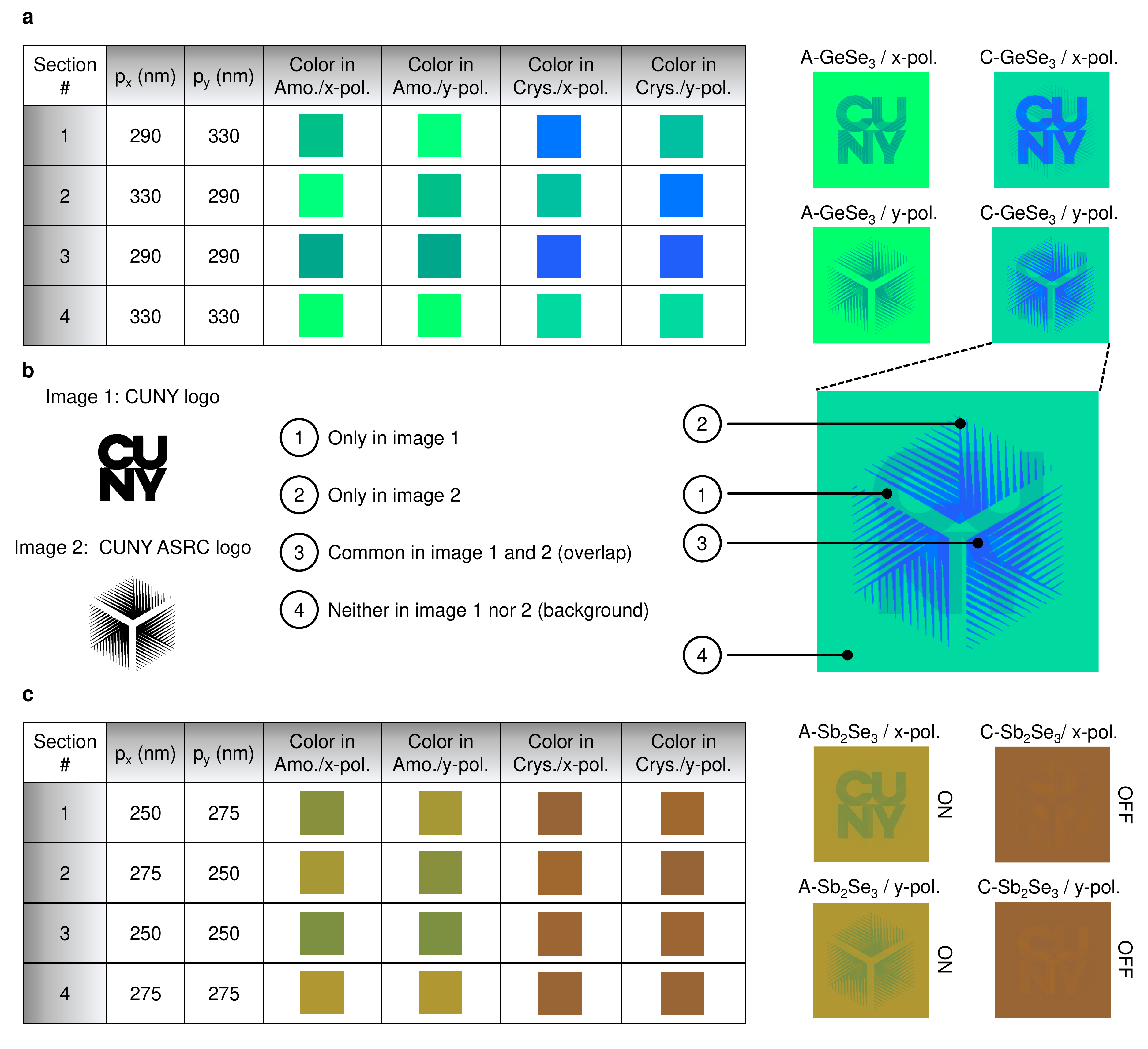}
\caption{\textbf{Design strategy for encryption of four different images into the phase and polarization of Sb$_2$Se$_3$ and GeSe$_3$ meta-pixels.} \textbf{a}, Polarization-based switching between two different images. The colors are generated by four different metasurfaces consisting of GeSe$_3$ nanopillars (Fig.~S3b) with periodicities reported in the table, diameters $d_{\textrm{x,y}} =0.55 \, p_{\textrm{x,y}} $, and a fixed height $h=250$ nm. \textbf{b}, The definition of different zones in each image. \textbf{c}, Phase-transition-based switching between the ON-state (amorphous) and the OFF-state (crystalline). The colors are generated by four different metasurfaces consisting of Sb$_2$Se$_3$ nanopillars with periodicities reported in the table, diameters $d_{\textrm{x,y}} =0.55 \, p_{\textrm{x,y}} $, and a fixed height $h=120$ nm.}
\label{Fig_S13}
\end{figure*}

\begin{figure*}[htbp]
\centering
\includegraphics[page=1,width=0.9\linewidth, trim={0cm 0cm 0cm 0cm},clip]{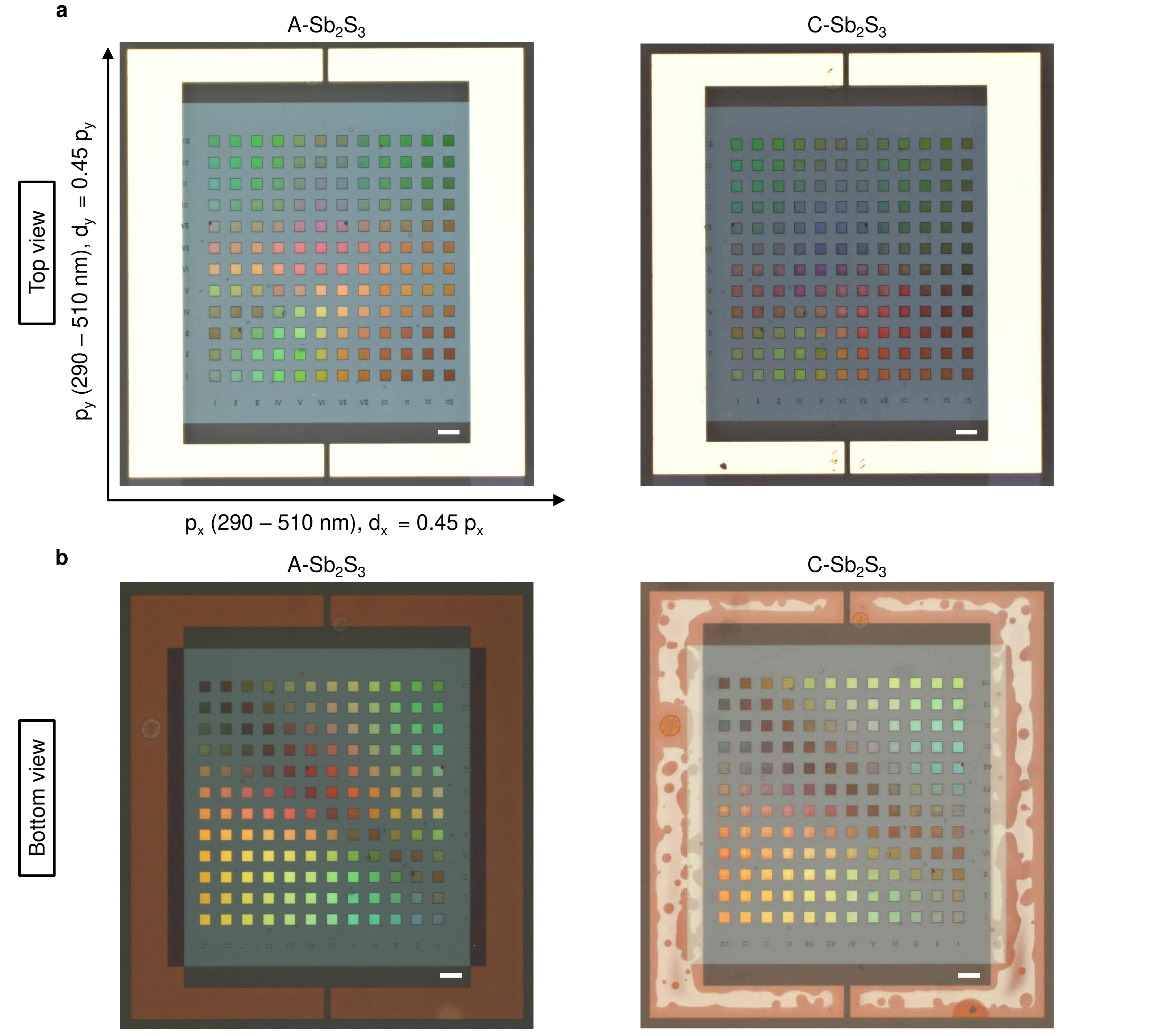}
\caption{\textbf{Electrical conversion of color palettes of Sb$_2$S$_3$ meta-pixels using ITO heater.} \textbf{a,b}, A-Sb$_2$S$_3$ (left) and C-Sb$_2$S$_3$ (right) meta-pixels observed from (\textbf{a}) top and (\textbf{b}) bottom of the sample considering different periodicities in x- and y- directions ($p_x$ and $p_y$, respectively) varying with 20 nm increments while the diameter of the Sb$_2$S$_3$ nanopillars changes as $d_{\textrm{x,y}} =0.45 \, p_{\textrm{x,y}}$, and the height of the nanopillars is fixed at $h=120$ nm. The scale bars are 100 $\mu$m.}
\label{Fig_S19}
\end{figure*}

\begin{figure*}[htbp]
\centering
\includegraphics[page=1,width=0.9\linewidth, trim={0cm 0cm 0cm 0cm},clip]{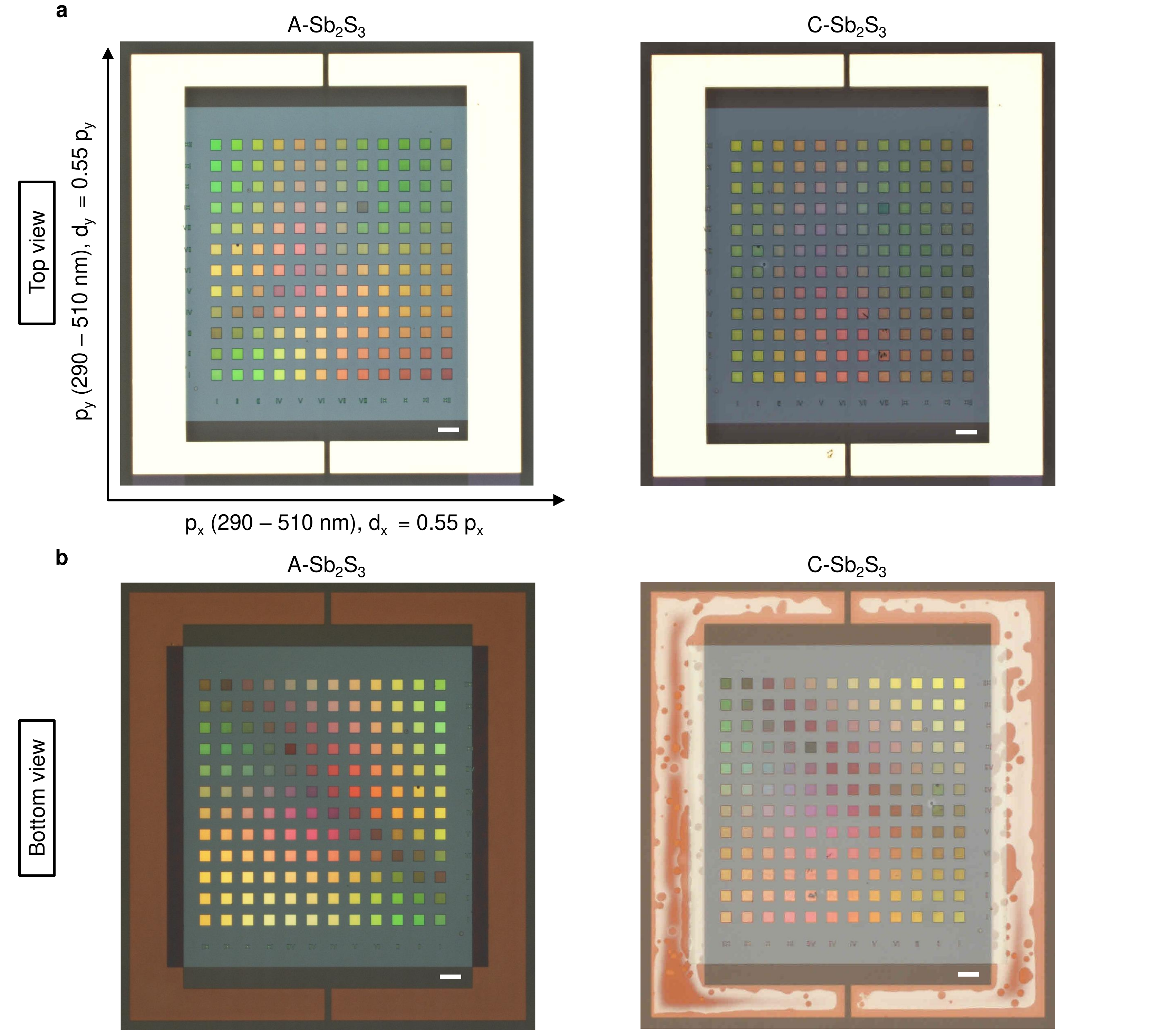}
\caption{\textbf{Electrical conversion of color palettes of Sb$_2$S$_3$ meta-pixels using ITO heater.} \textbf{a,b}, A-Sb$_2$S$_3$ (left) and C-Sb$_2$S$_3$ (right) meta-pixels observed from (\textbf{a}) top and (\textbf{b}) bottom of the sample considering different periodicities in x- and y- directions ($p_x$ and $p_y$, respectively) varying with 20 nm increments while the diameter of the Sb$_2$S$_3$ nanopillars changes as $d_{\textrm{x,y}} =0.55 \, p_{\textrm{x,y}}$, and the height of the nanopillars is fixed at $h=120$ nm. The scale bars are 100 $\mu$m.}
\label{Fig_S20}
\end{figure*}

\begin{figure*}[htbp]
\centering
\includegraphics[page=1,width=0.9\linewidth, trim={0cm 0cm 0cm 0cm},clip]{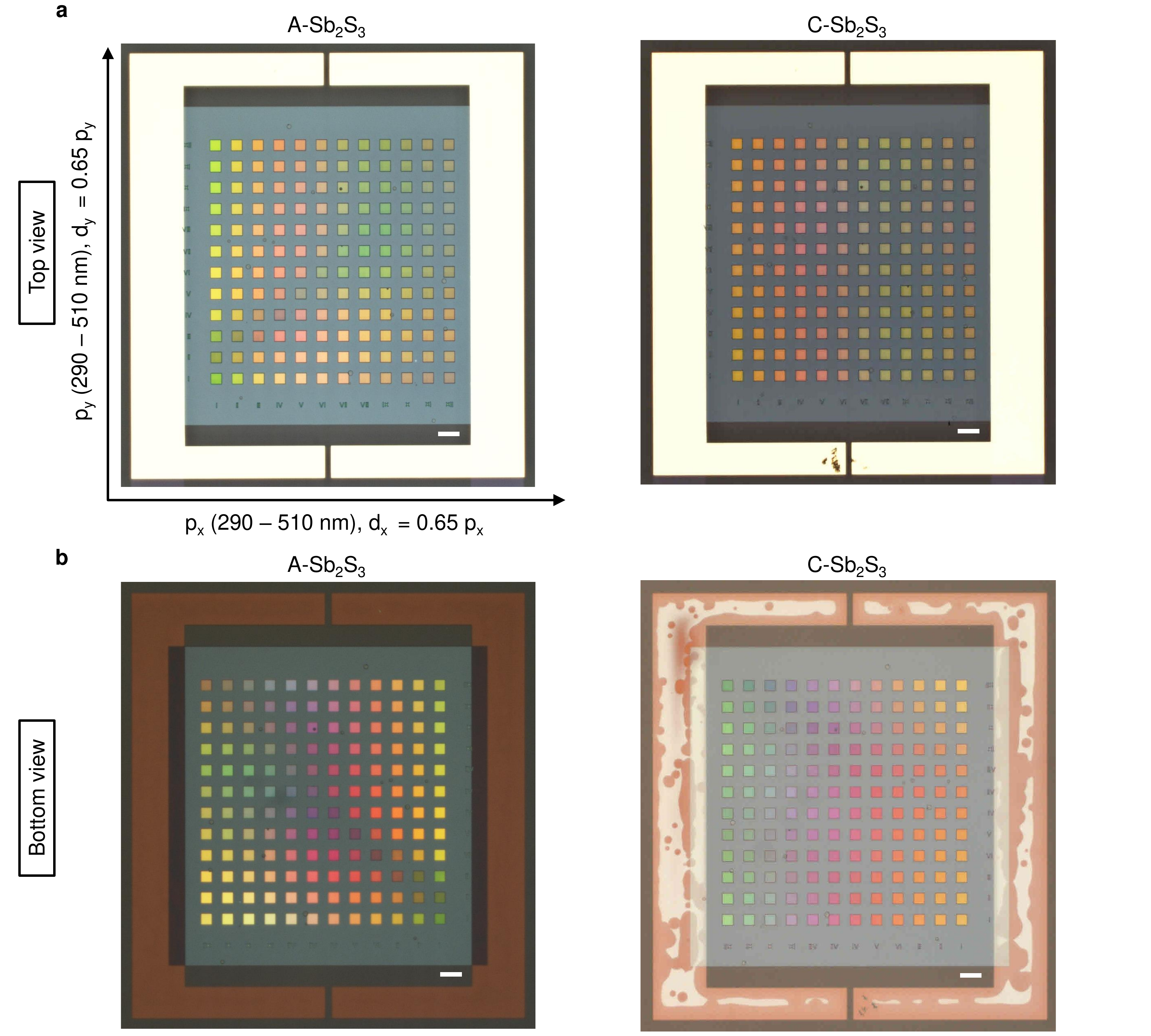}
\caption{\textbf{Electrical conversion of color palettes of Sb$_2$S$_3$ meta-pixels using ITO heater.} \textbf{a,b}, A-Sb$_2$S$_3$ (left) and C-Sb$_2$S$_3$ (right) meta-pixels observed from (\textbf{a}) top and (\textbf{b}) bottom of the sample considering different periodicities in x- and y- directions ($p_x$ and $p_y$, respectively) varying with 20 nm increments while the diameter of the Sb$_2$S$_3$ nanopillars changes as $d_{\textrm{x,y}} =0.65 \, p_{\textrm{x,y}}$, and the height of the nanopillars is fixed at $h=120$ nm. The scale bars are 100 $\mu$m.}
\label{Fig_S21}
\end{figure*}

\begin{figure*}[htbp]
\centering
\includegraphics[page=1,width=0.9\linewidth, trim={0cm 0cm 0cm 0cm},clip]{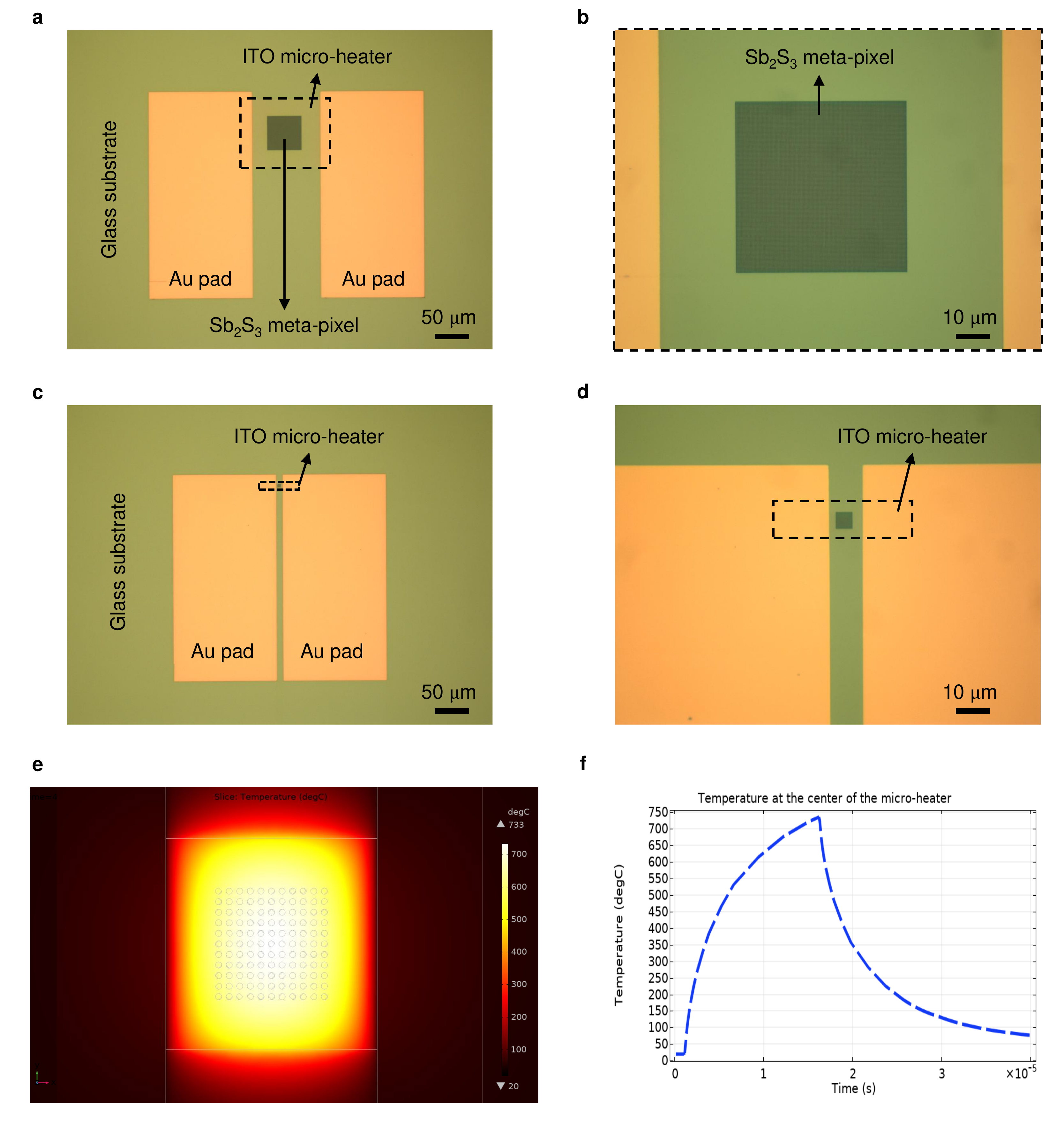}
\caption{\textbf{Electrical conversion of a Sb$_2$S$_3$ meta-pixel using ITO micro-heater.} \textbf{a-d}, Microscope images of (\textbf{a,b}) 100$\times$100 $\mu$m$^2$, and (\textbf{c,d}) 10$\times$10 $\mu$m$^2$ micro-heaters with 50$\times$50 $\mu$m$^2$ and 5$\times$5 $\mu$m$^2$ meta-pixels at the center, respectively. \textbf{e}, Simulated temperature distribution in the cross-section of the meta-pixel in (\textbf{d}) at the end of a 7 V pulse with 15 $\mu$s duration. \textbf{d}, Real-time temperature profile at the center of the meta-pixel upon applying the re-amorphization pulse to the microheater.}
\label{Fig_S22}
\end{figure*}

\begin{figure*}[htbp]
\centering
\includegraphics[page=1,width=0.9\linewidth, trim={0cm 0cm 0cm 0cm},clip]{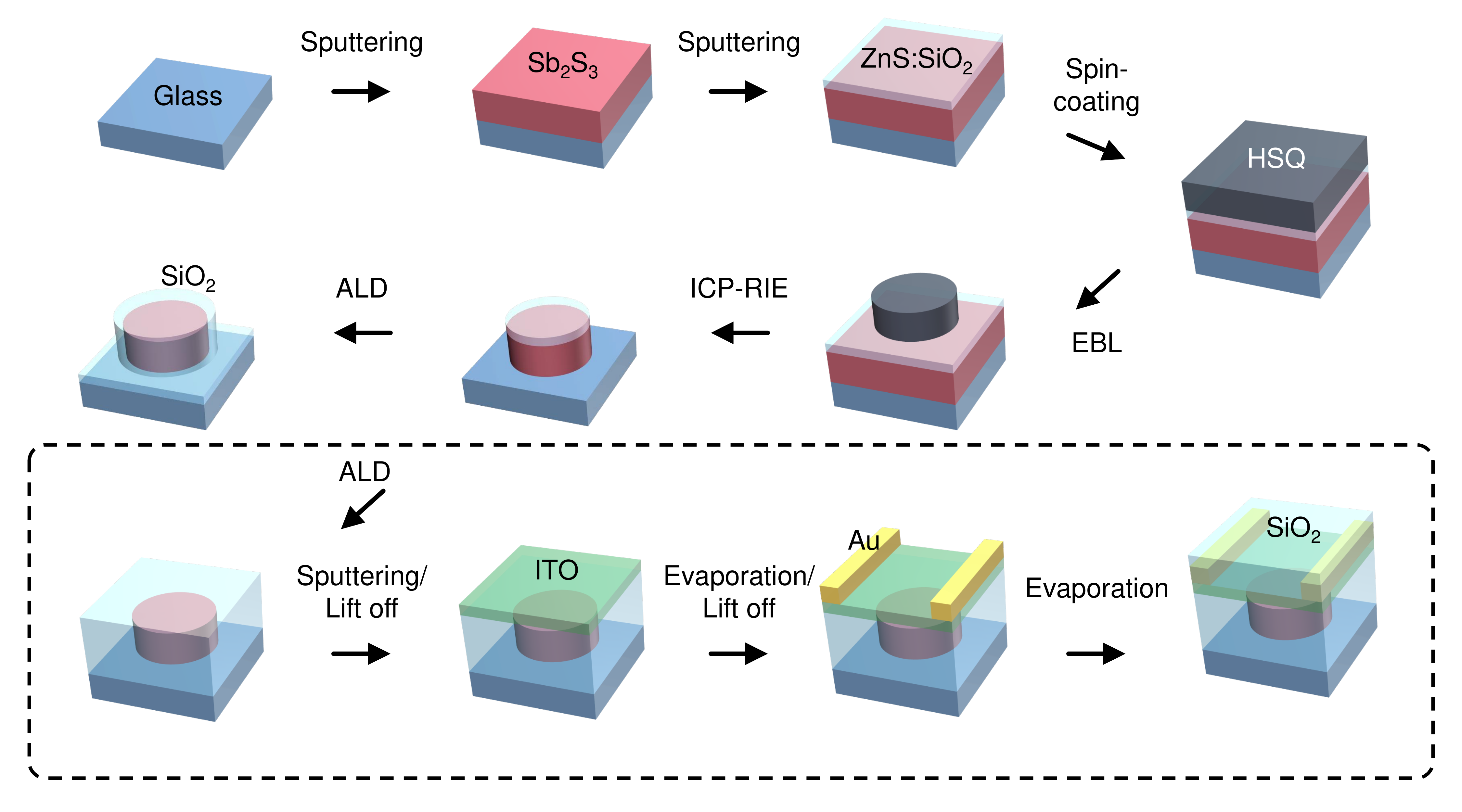}
\caption{\textbf{Fabrication process.}}
\label{Fig_S23}
\end{figure*}

\begin{figure*}[htbp]
\centering
\includegraphics[page=1,width=1\linewidth, trim={0cm 0cm 0cm 0cm},clip]{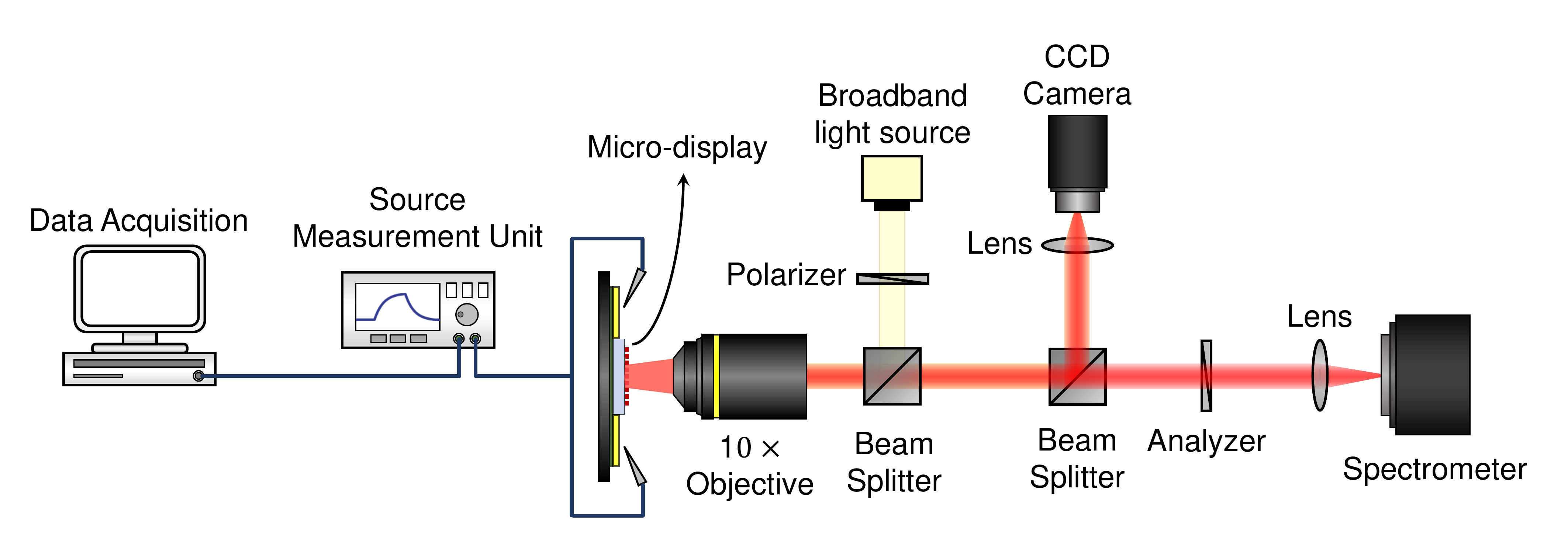}
\caption{\textbf{Optical characterization setup.}}
\label{Fig_S24}
\end{figure*}

\begin{figure*}[htbp]
\centering
\includegraphics[page=1,width=.7\linewidth, trim={0cm 0cm 0cm 0cm},clip]{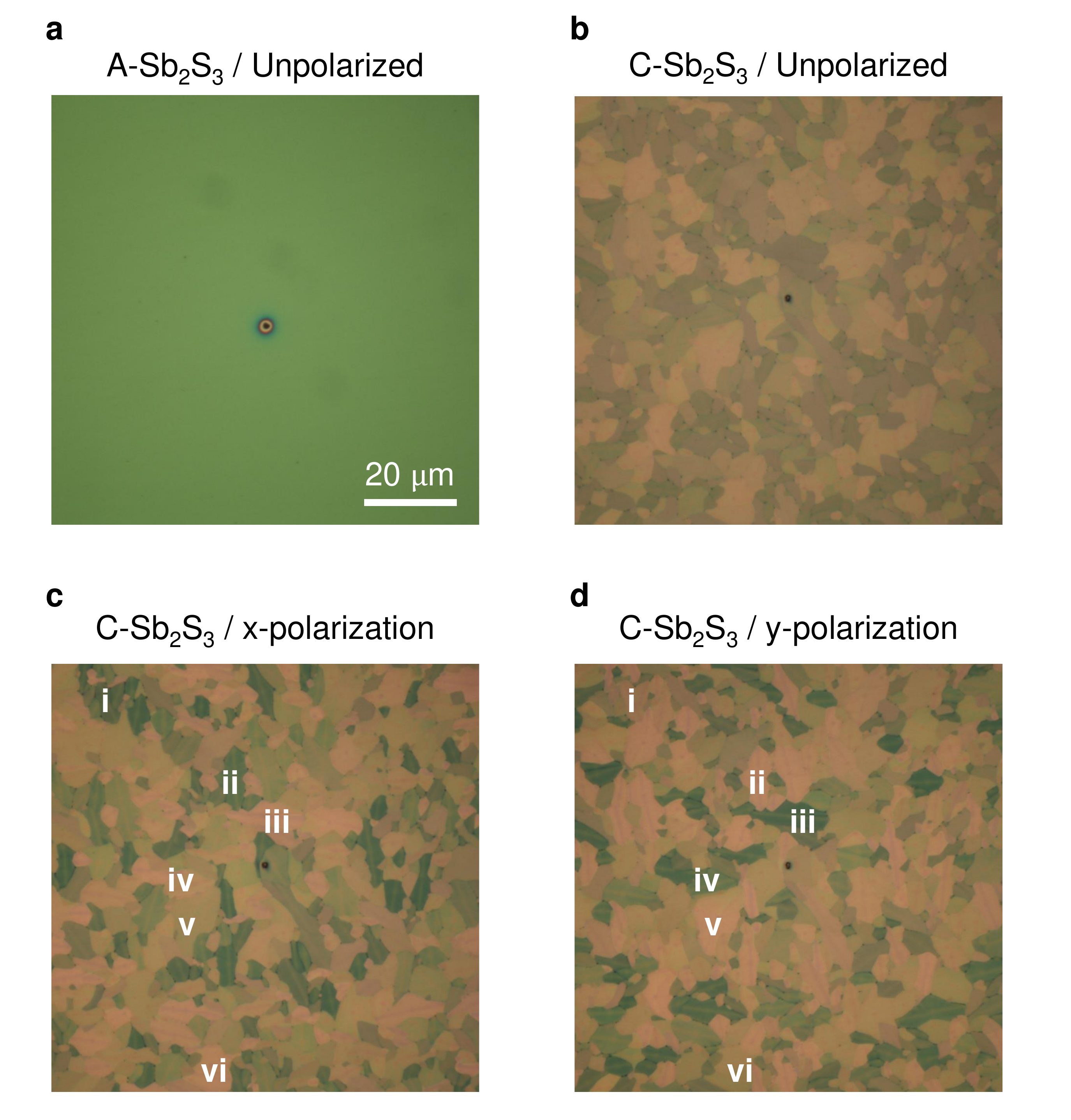}
\caption{\textbf{Characterization of the anisotropic C-Sb$_2$S$_3$ crystals.} \textbf{a-d}, Optical images a film of (\textbf{a}) A-Sb$_2$S$_3$ and (\textbf{b-d}) C-Sb$_2$S$_3$ under microscope with (\textbf{b}) unpolarized, (\textbf{c}) x- and (\textbf{d}) y-polarized incident white light. The crystalized regions at (i) and (ii) switches from greenish colors to brownish ones going from x- to y-polarization, while colors in regions (iii) and (iv) changes from brownish to greenish, and colors in areas (v) and (vi) remains the almost unchanged. The dark particle at the center of the images is used as the marker for positioning.}
\label{Fig_S25}
\end{figure*}

\begin{figure*}[htbp]
\centering
\includegraphics[page=1,width=.7\linewidth, trim={0cm 0cm 0cm 0cm},clip]{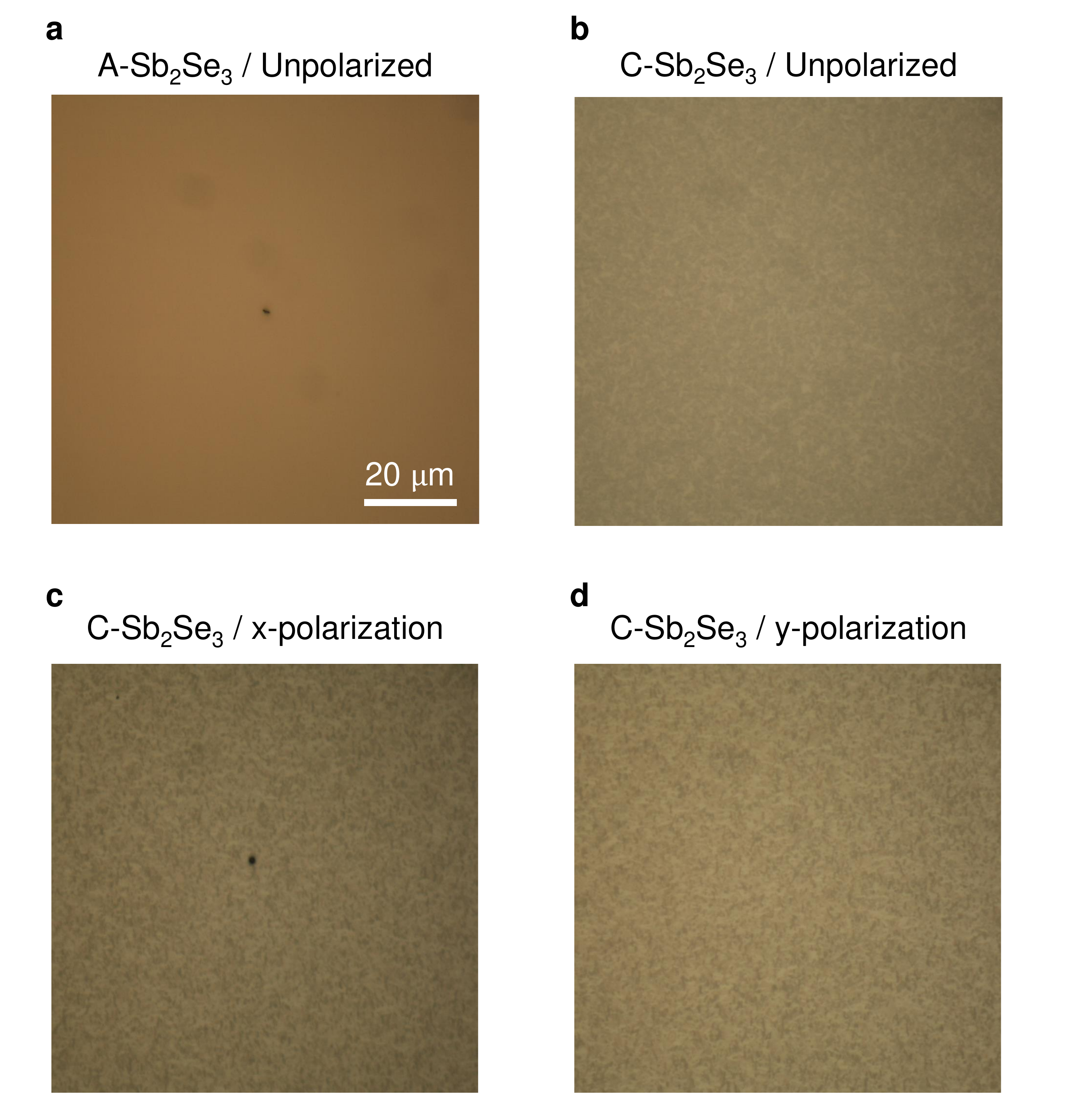}
\caption{\textbf{Characterization of the anisotropic C-Sb$_2$Se$_3$ crystals.} \textbf{a-d}, Optical images a film of (\textbf{a}) A-Sb$_2$Se$_3$ and (\textbf{b-d}) C-Sb$_2$Se$_3$ under microscope with (\textbf{b}) unpolarized, (\textbf{c}) x- and (\textbf{d}) y-polarized incident white light. The dark particle at the center of the images is used as the marker for positioning.}
\label{Fig_S26}
\end{figure*}



\newpage
\newpage

\end{document}